\documentclass[twocolumn,preprintnumbers,amsmath,amssymb,aip]{revtex4}

\usepackage{graphicx}
\usepackage{dcolumn}
\usepackage{bm}
\usepackage{xcolor}
\usepackage{natbib}
\begin{document}

\title{Universal scalings for laser acceleration  of electrons in ion channels}

\author{ Vladimir Khudik, Alexey Arefiev, Xi Zhang, and Gennady Shvets}

\affiliation{Department of Physics and Institute for Fusion Studies, The University of Texas at Austin, Austin, Texas 78712, USA}
\date{\today}
\begin{abstract}
   Direct laser acceleration  of  electrons in ion channels is investigated in a general case when the laser phase velocity is greater than (or equal to) the speed  of  light.
	Using the similarity of   the equations of motion for ultra-relativistic electrons, we develop a universal scaling theory that  
	gives  the maximum possible energy that can be attained by an electron for given laser and plasma parameters. The theory predicts
	 appearance of  forbidden zones in the phase space of the particle, which manifests itself  as an energy gain threshold.  
	We apply the developed theory to find the conditions  needed for  an energy enhancement via a resonant interaction between the third harmonic of betatron oscillations and the laser wave. The theory is also used to analyze electron dynamics in a circularly polarized laser.
\end{abstract}
 
\maketitle


%
%
 



\section{Introduction}
High intensity  laser beams  propagating in an underdense plasma expel  ambient plasma electrons  radially and create  a zone fully or partially evacuated of electrons. Short beams  generate a structure, known as a plasma bubble, that follows in the wake of the laser pulse~\cite{pukhov_bubble},  whereas long beams create slowly evolving quasistatic ion channels~\cite{pukhov_channel}. In the bubble regime,  there is a strong longitudinal electric field that can  accelerate   electrons to very high energies while a transverse electromagnetic field  keeps electrons  close to the bubble axis~\cite{leemans_1gev, wang_2gev, leemans_4gev, kim_3gev}. If a second laser pulse is placed close to the  bottom of the plasma bubble, then the electrons can undergo direct laser acceleration (DLA) in this  pulse and gain  significant energy~\cite{zhang_prl,zhang_ppcf}.  In the ion channel regime, the longitudinal field is usually  weak and electrons are accelerated  only by the laser. This regime is interesting for applications that require a large number of  energetic electrons.   Specifically, generation of copious relativistic electrons is the key to x-ray~\cite{Park2006,Kneip2008} and secondary particle sources, such as energetic ions~\cite{Schollmeier2015}, neutrons~\cite{Pomerantz2014}, and positrons~\cite{Chen2015}.




Direct laser acceleration (DLA) of electrons in  ion channels and plasma bubbles has been considered analytically and through extensive computational work~\cite{shaw_ppcf, zhang_prl,zhang_ppcf, suk_pop}. Recently it has been shown   that due to  parametric instability~\cite{Arefiev2012, Arefiev2014}  the electron trajectory can become essentially  a  three dimensional curve even when electrons are injected in  the plane formed by the channel axes and the laser polarization~\cite{Arefiev2016}.  


Transverse electric fields of the channel can significantly alter electron oscillations across the channel during direct laser acceleration. This can allow for a  resonant interaction where the electron transverse velocity remains anti-parallel to the laser electric field over extended segments of the electron trajectory.  During the resonant interaction, the electron can gain significant energy from the laser electric field, which is then converted into the longitudinal motion by the laser magnetic field.  
 Distinctive  feature of the resonant interaction of electrons with  high intensity laser wave  is  that the Doppler-shifted frequency of the  wave  can  oscillate from almost zero  to its maximum value while the betatron frequency of  transverse electron oscillations  experiences relatively slow variations~\cite{Shaw_2015}. Such a nontrivial resonant interaction complicates the electron dynamics and makes the specifics of interaction mechanism  unclear.


In this paper  we develop an  analytical description of the  acceleration of electrons by linearly polarized laser wave in  ion channels and obtain the universal scalings for the maximum electron energy.
The paper is organized as follows. In Sec.~II, we discuss the paraxial approximation of  equations of motion for electrons accelerated by the laser wave and show that in dimensionless variables these equations depend only on two parameters. In Sec.~III, we  develop main components  of universal scalings theory for the luminal laser wave: averaging  equations of motion over betatron oscillations~\cite{Davoine2014,Whittum1992}, finding  their analytical solution, and explaining how appearance of the forbidden zones in the phase space results in thresholds in dependence of maximum electron energy on the laser-plasma parameters.  In Sec.~IV, the similar consideration is developed for super-luminal laser wave.
Then in Sec.~V, we consider several spinoffs of the universal scalings theory: (A) acceleration of electrons through the  resonance between the wave and the third harmonic of betatron oscillations, (B) acceleration of electrons by the circularly polarized laser wave, and (C) acceleration of pre-accelerated electrons. Finally, we summarized and discussed the obtained results in Sec.~VI.  

\section{Electron motion in the paraxial approximation} 
We examine  electron dynamics in the framework of a  model that incorporates a  laser beam  propagating along the axis of a uniform  cylindrical ion channel and a focusing electrostatic field created by the  channel ions. The fields of the planar linearly polarized  laser pulse are  
$E_y^{(L)} = E_0 \cos{\phi}$ and $B_z^{(L)} = E_y^{(L)}c/v_{ph}$
where $\phi=\omega_L (x/v_{\rm ph} - t)$ is the wave phase, $E_0=a_0(m\omega_Lc/e)$ is the amplitude of the laser electric  field. The electrostatic field created by the channel ions is approximated by ${\bf{E}}_{\perp} =  m\omega_p^2 {\bf{r}}_{\perp}/2e$ where $\omega_p = \sqrt{4\pi e^2 n/m}$ is the plasma frequency, $n$ is the  density of the uncompensated positive charge, and $m$ is the electron mass and subscript '$\perp$' denotes the vector components transverse to the channel axis. The equations of motion are then given by
\begin{eqnarray}
 \frac{d p_x}{dt} =-\frac{e}{c}v_y B_z^{(L)}, \label{eq:EqM_N1}\\
\frac{d {\bf{p}}_{\perp}}{dt} = -\frac{1}{2}m_e\omega_p^2 {\bf{r}}_{\perp}+e\Big(\frac{v_x}{c}B_z^{(L)}-E_y^{(L)}\Big){\bf{e}}_y, \label{eq:EqM_N2}\\
\frac{d {\bf{r}}}{dt}={\bf{v}}=\frac{\bf{p}}{m\gamma},
\label{eq:EqM_N3}
\end{eqnarray} 
where  ${\bf{e}}_y$ is the unit vector directed along the $y$-axis, and $\gamma=(1+{\bf{p}}^2/m^2c^2)^{1/2}$ is the electron relativistic factor.

The electron dynamics in the ion channel  is determined to a large extent by  two   frequencies pertinent to  Eqs.~(\ref{eq:EqM_N1}) - (\ref{eq:EqM_N3}): the frequency of  natural oscillations of the electron across the channel (betatron frequency) and the  frequency of the oscillating laser fields experienced by the moving electron (Doppler shifted frequency): 
\begin{eqnarray}
  \omega_{\beta}=\omega_p/(2\gamma)^{1/2},\label{omega_b}\\
\omega_D=\omega_L(1-v_x/v_{ph}).\label{omega_D}
\end{eqnarray}
Note that, by definition, we have $d\phi/dt=-\omega_D$. 

  One can check that during particle motion the  'energy' in the co-moving coordinates $\xi=x-v_{ph}t$ is conserved 
\begin{eqnarray}
 \gamma - \frac{v_{ph}}{c}\frac{p_x}{m_e c} + \frac{\omega_{p}^2}{4 c^2} {\bf{r}}_{\perp}^2 = \mathcal{I}_0=\mbox{const}, \label{R_def}
\end{eqnarray}
where  $\mathcal{I}_0$ is a constant determined from initial conditions.

It is well known~\cite{Stupakov2001} that when a relativistic intensity electromagnetic wave (with $a_0\gg 1$) accelerates electrons in vacuum, the typical value of the electron momentum is $p_x\sim m_ec a_0^2/2\gg p_y \sim  mc a_0\gg mc$, so that $\gamma\approx p_xc$. Similarly, we expect for the energy of ultra-relativistic electrons accelerated in the ion channel to be primarily associated with the momentum in the direction of the wave propagation. Under the paraxial approximation ($p_x>>|{\bf{p}}_{\perp}|>>m_ec$), equations~(\ref{eq:EqM_N3}) reduce to
\begin{eqnarray}
\frac{d x}{dt}=v_x=c\Big(1-\frac{{\bf{v}}_{\perp}^2}{2c^2}\Big), \quad \frac{d {\bf{r}}_{\perp}}{dt}={\bf{v}}_{\perp}=c\frac{{\bf{p}}_{\perp}}{p_x}.  \label{eq:APPROX}
\end{eqnarray}
The integral of motion~(\ref{R_def}) relates   the transverse energy of the particle  $\epsilon_{\perp}$ to its longitudinal momentum $p_x$:  
\begin{eqnarray}
 \epsilon_{\perp}=\mathcal{I}_0mc^2+({v_{ph}}-{c}){p_x}, \quad  \epsilon_{\perp}\equiv \frac{p_x{\bf{v}}_{\perp}^2}{2c}+\frac{1}{4}\omega_p^2m{\bf{r}}_{\perp}^2.
\label{eq:Energy}
\end{eqnarray}
 It follows from this conservation law that the characteristic amplitude of transverse oscillations is given by $ r_*\equiv 2\mathcal{I}_0^{1/2}c/\omega_p$. The characteristic longitudinal momentum $p_*$ during electron acceleration can be found from the condition that the frequency of betatron oscillations  matches the Doppler shifted frequency: $ \omega_p/(2p_x/mc)^{1/2}\approx \omega_L (1-v_x/v_{ph})$. Using Eqs.~(\ref{eq:APPROX}) and (\ref{eq:Energy}), we find that   $p_*\sim mc \omega_L^2/\omega_p^2$ and that the characteristic frequency is $\omega_*\sim \omega_p^2/\omega_L$.

 These  estimates suggest the introduction of new dimensionless variables
\begin{eqnarray}
\tilde{t}=\Omega_*\omega_p t,\quad 
\tilde{{\bf{r}}}_{\perp}=\frac{{{\bf{r}}}_{\perp}}{r_*},\quad
\tilde{p}_x=\frac{p_x}{p_*}\quad
\tilde{{\bf{p}}}_{\perp}=\frac{{{\bf{p}}}_{\perp}}{p_{\perp*}},\\
 {\Omega_*}\equiv \frac{\tilde{v}_{ph}\omega_p}{\mathcal{I}_0\omega_L },\,\,\,
{p_*}\equiv \frac{{mc}}{{2\Omega_*^2}},\,\,\, {p_{\perp*}}\equiv\frac{mc \mathcal{I}_0^{1/2}}{\Omega_*},
 \end{eqnarray} 
where  $\tilde{v}_{ph}\equiv {v_{ph}}/{c}$ and $\omega_*\equiv\Omega_*\omega_p$. Using  the phase of the   wave $\phi=\omega_L(x/v_{ph}-t)$ instead of the coordinate $x$, we can transform Eqs.~(\ref{eq:EqM_N1}), (\ref{eq:EqM_N2}), and (\ref{eq:APPROX})  to the following dimensionless form
\begin{eqnarray}
 \dot{\tilde{p}}_x =-4\mathcal{E}\dot{\tilde{y}} \cos{\phi},\quad
{\dot{\tilde{\bf{p}}}}_{\perp}=-{\tilde{\bf{r}}}_{\perp}+{\bf{e}}_y\dot{\phi}\mathcal{E} \cos \phi,\label{eq:Eq_AA_2}\\
 \dot{\phi} =-2(\dot{\tilde{{\bf{r}}}}_{\perp}^2+\chi),\quad 
\dot{{\tilde{{\bf{r}}}}}_{\perp}={\tilde{\bf{p}}}_{\perp}/\tilde{p}_x,  \label{eq:Eq_AA_3}
  \end{eqnarray} 
where the 'dot' above the variable denotes a derivative with respect to the dimensionless time $\tilde{t}$. 
 The parameters  $\mathcal{E}$ and $\chi$ characterize the strength of the laser wave and  its dispersion  in the ion channel: 
\begin{eqnarray}
\mathcal{E}\equiv a_0({\omega_p}/{\omega_L}){\tilde{v}_{ph}}{\mathcal{I}_0^{-3/2}},\\ \chi\equiv ({\mathcal{I}_0}/{\tilde{v}_{ph}^2}){(\tilde{v}_{ph}-1)}/({2\omega_p^2/\omega_L^2})
\end{eqnarray} 
The integral of motion in dimensionless variables takes the form:
\begin{eqnarray}
\tilde{p}_x\dot{\tilde{{\bf{r}}}}_{\perp}^2+\tilde{{\bf{r}}}_{\perp}^2=1+\chi \tilde{p}_x,
  \label{eq:ENERGY_PARAX}
\end{eqnarray} 
 
Initially sub-relativistic electrons perform irregular oscillations and their dynamics is  determined by several parameters $\omega_p$. However, once the 
electron energy becomes  ultra-relativistic, the electron motion can be described by the  the paraxial approximation (\ref{eq:Eq_AA_2}) and (\ref{eq:Eq_AA_3}) which contains only  two parameters $\mathcal{E}$ and $\chi$. Since  electrons gain most of their  energy  during this stage,  
it raises the question in what extent the electron dynamics is determined by  these  parameters.

\begin{figure}
	\centering
   	\includegraphics[height=0.11\textheight,width=0.85\columnwidth]{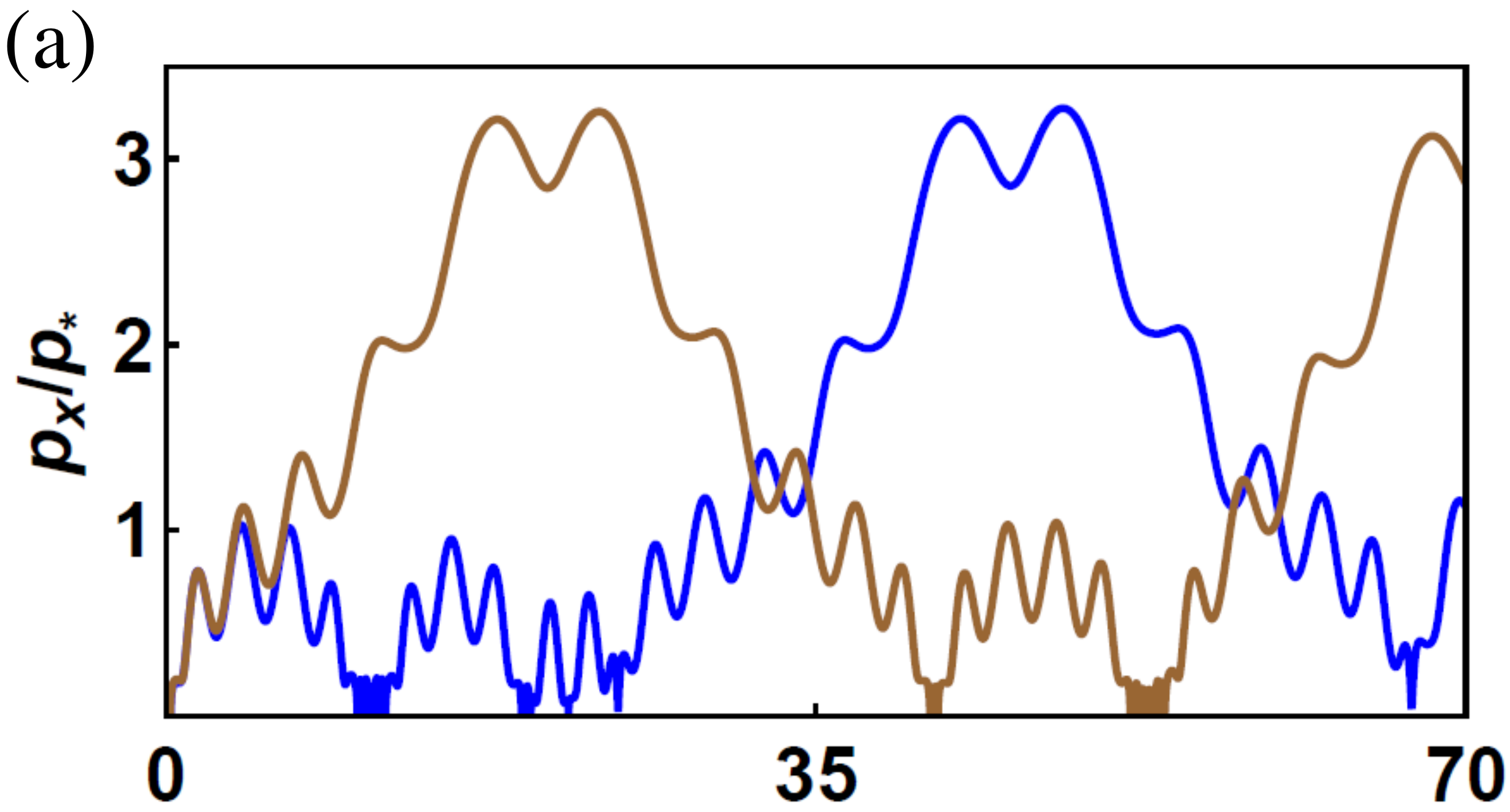}  \\
	\hspace{3mm}\includegraphics[height=0.13\textheight,width=0.87\columnwidth]{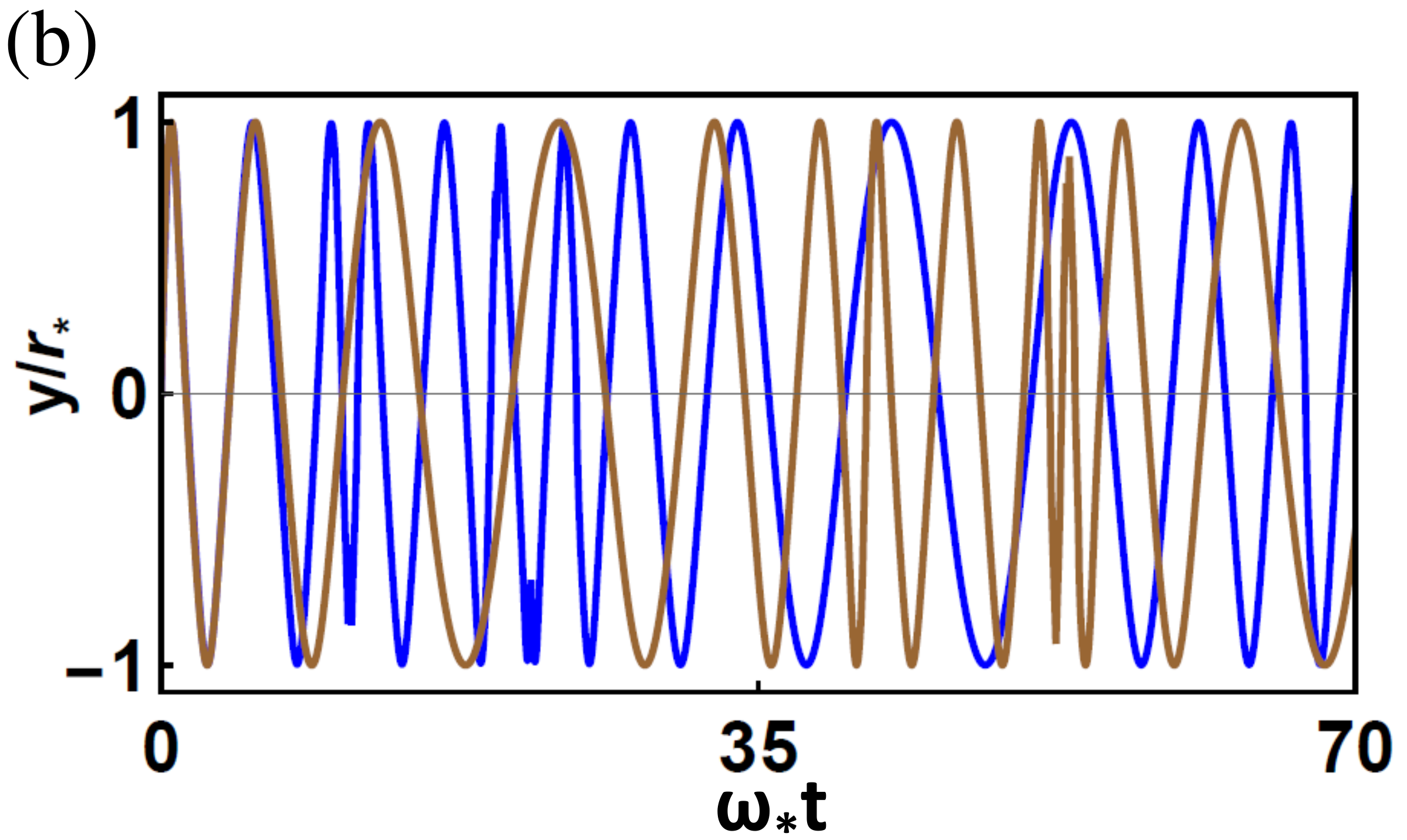} 
  \caption{
	 Dynamics of  electrons  initially placed at rest on the channel axis at $\phi|_{t=0}=\frac{\pi}{2}$ for parameters $\mathcal{E}=0.2$ and $\chi=0$ ($v_{ph}=c$).  Blue lines correspond to the electron acceleration in the case when  $\omega_p/\omega_L = 0.05$ and   $a_0 =4$ ($p_*=200$ and $\omega_*=\omega_L/400$), and  brown lines - in the case when $\omega_p/\omega = 0.02$ and  $a_0 =10$ ($p_*=1250$ and $\omega_*=\omega_L/2500$). }\label{Figure1_1}
\end{figure}

In all examples below we assume for simplicity that initially electrons are placed at rest on the axis of the channel, so that the only  difference  in their initial conditions is the logitudinal location given by  $x$ or, equivalently,  by the wave phase $\phi$.  Figure~\ref{Figure1_1} shows the result of integration of exact equations of motion~(\ref{eq:EqM_N1}) - (\ref{eq:EqM_N3}) for two electrons accelerated by the laser in the ion channel  with the same parameters $\mathcal{E}=0.2$ and $\chi=0$ but with  different $a_0$ and $\omega_p/\omega_L$. In both cases, the electrons  initially perform irregular oscillations and then they gain a large energy following somewhat similar peaks, see Fig.~\ref{Figure1_1} (a). The peaks with the largest energy gain can differ  in details but always exhibit a general  similarity. 
 Figure~\ref{Figure1_1} (b) shows that in the luminal case the  amplitude of betatron oscillations  is constant and equal to $r_*$ once the   electrons have gained ultra-relativistic energies.     

\section{Scalings in the luminal case }

In this section we evaluate the maximum energy that can be gained from the laser at given  laser and plasma parameters. We show that small changes of the parameter $\mathcal{E}$ can result in dramatic changes in the electron motion. Specifically, this aspect manifests itself as  threshold dependence of the maximum energy gain on the laser amplitude. 

\subsection{Analytical theory} \label{SecA}

In the paraxial approximation   the dimensionless amplitude of  oscillations  is equal to unity and thus it is convenient to introduce a phase of betatron oscillations $\psi$, such that  $\tilde{y}=\sin{\psi}$.  Using this relashionship, we immediately find from Eq.~(\ref{eq:ENERGY_PARAX}) by setting  $\chi=0$   that $\dot{\tilde{y}}=\cos\psi/{\tilde{p}_x}^{1/2}$ and $\dot{\psi}=1/{\tilde{p}_x}^{1/2}$. The latter expression relates the change of the phase of transverse oscillations to the betatron frequency $\dot{\psi}= \omega_{\beta}/\omega_*$ [see Eq.~(\ref{omega_b})].

To make further progress, we  utilize the other  important feature evident from Fig.~\ref{Figure1_1}, namely, that the particles perform several (and in some cases many) betatron oscillations while  gaining their energy. Averaging out these oscillations  significantly simplifies the description of the ultra-relativistic motion []. Equation~(\ref{eq:Eq_AA_3}) for the wave phase then reduces to $\langle\dot{\phi}\rangle =-2\langle{\dot{\tilde{y}}^2}\rangle\approx -1/\tilde{p}_x$, where the angle brackets denote  averaging over the betatron period.  Here we have neglected the difference between $\langle \tilde{p}_x \rangle$ and $\tilde{p}_x$ because of the slow change of the longitudinal momentum during the energy gain. 

As already pointed out, an effective energy gain requires for   the electron transverse velocity to remain anti-parallel to the laser electric field over extended segments of the electron trajectory.  
Therefore, the phase shift between the  phase of the laser wave and the phase of the  transverse oscillations, $\theta=\psi+\phi$, is an important characteristic of the laser-particle interaction.   
Combining the expressions for $\dot{\psi}$ and  $\dot{\phi}$, we obtain:
 \begin{eqnarray}
 \langle\dot{\theta}\rangle =
\frac{1}{{\tilde{p}_x}^{1/2}}-\frac{1}{{\tilde{p}_x}}.  \label{eq:Eq_theta}
\end{eqnarray} 
 To make clearer the physical meaning of the terms in the right hand  side of this equation we rewrite it in the following equivalent form: $\langle\dot{\theta}\rangle =
({\omega_{\beta}-\langle \omega_D\rangle})/{\omega_*}$ where $\omega_{\beta}=\omega_p/(2p_x/mc)^{1/2}$ and $\langle \omega_D\rangle=\omega_L\mathcal{I}_0/(2p_x/mc)$.  Note that $\omega_{\beta}=\langle \omega_D\rangle=\omega_*$  at $p_x=p_*$.

\begin{figure}[t]
 \centering
\includegraphics[height=0.13\textheight,width=0.85\columnwidth]{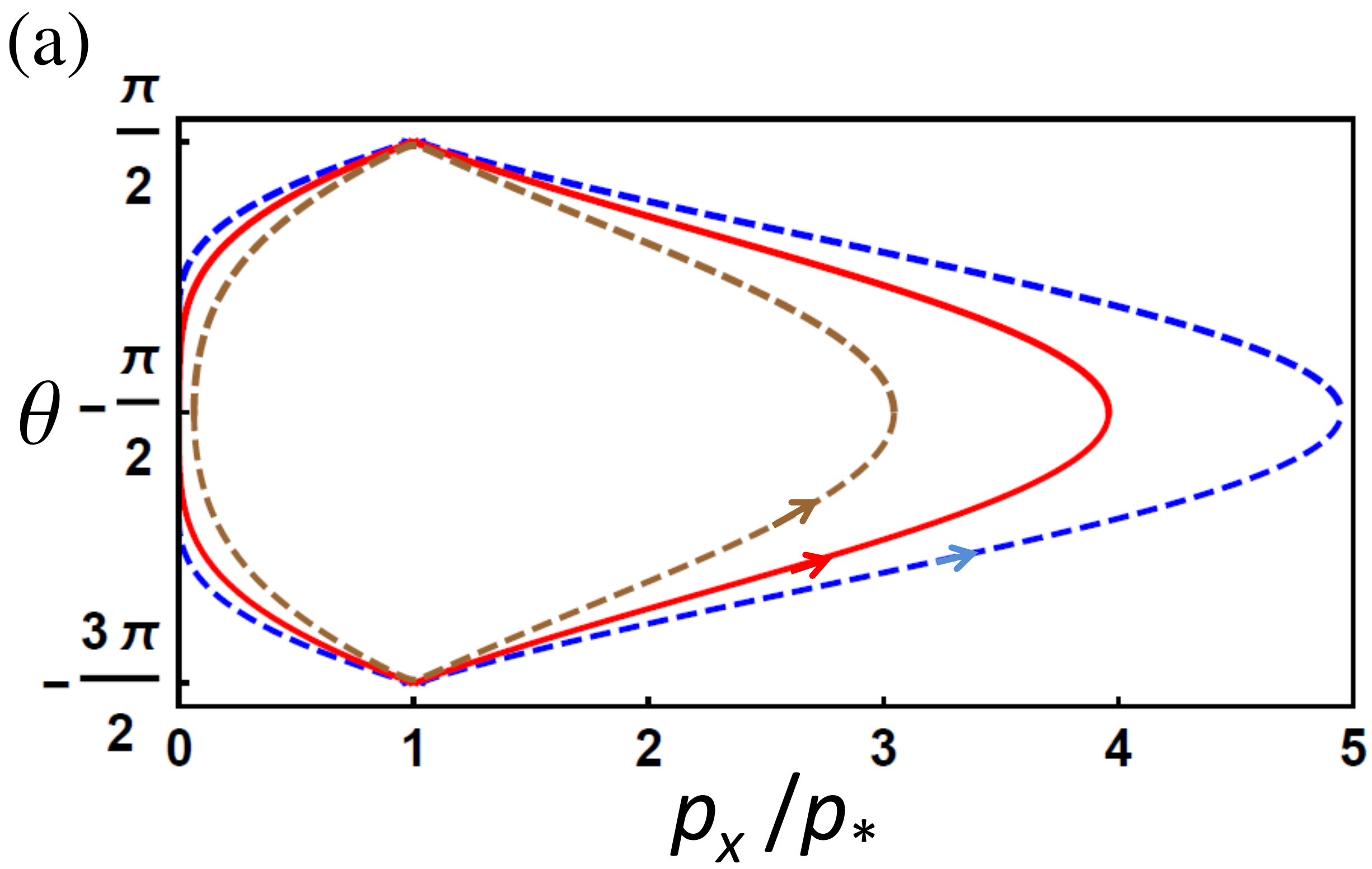}\\
\includegraphics[height=0.13\textheight,width=0.85\columnwidth]{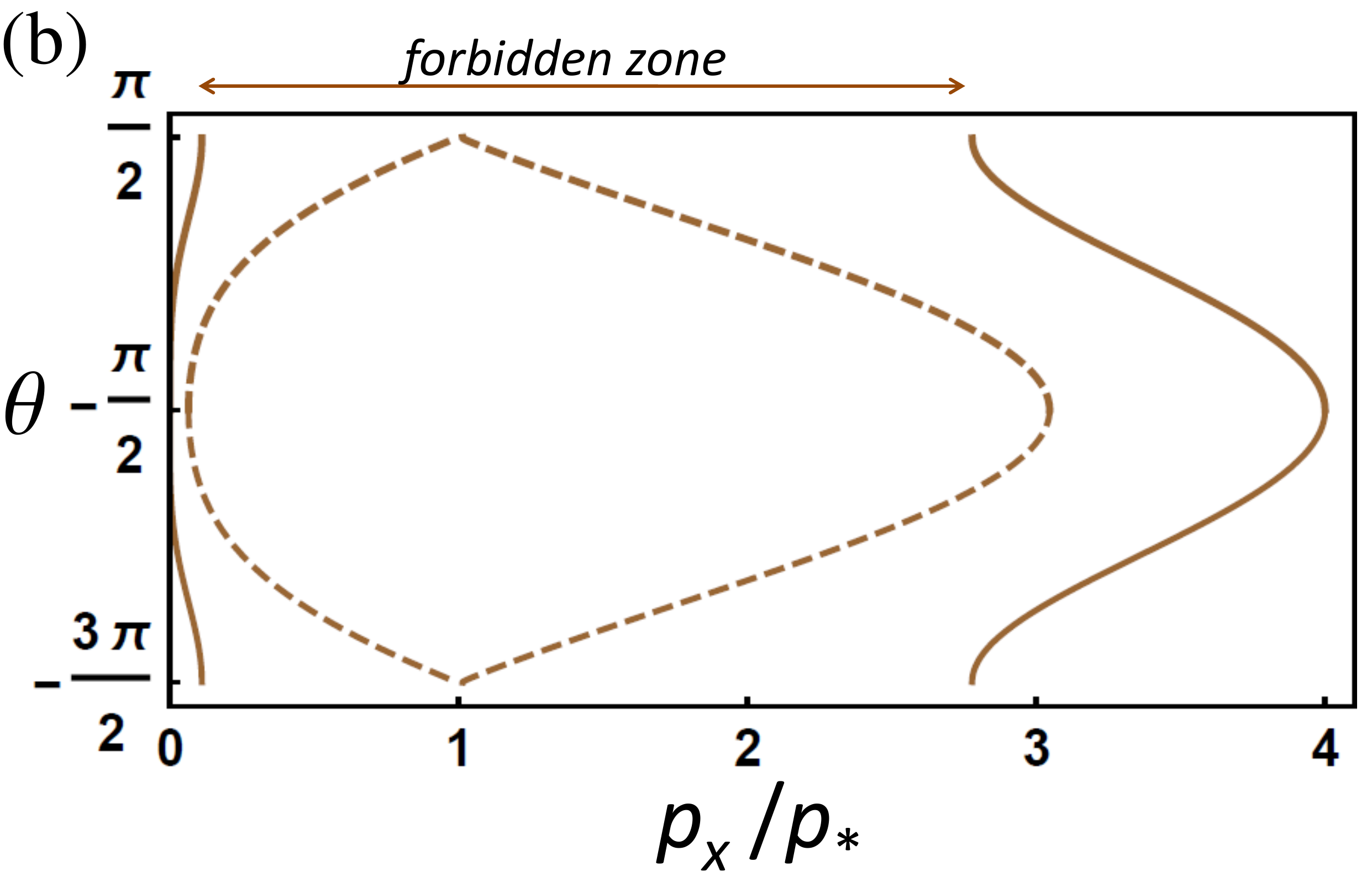}
\caption{ (a) Phase trajectories averaged over betatron oscillations moving along which electrons reach maximum energy: $\mathcal{E}=0.54$ (blue dashed), $\mathcal{E}=1/(8\alpha)\approx 0.36$ (red) and $\mathcal{E}=0.2$ (brown); in the latter case the initial (minimal) electron energy is greater than zero. (b) Forbidden zone in the phase space $(\tilde{p}_x,\theta)$ for   initially cold electrons moving  at sub-critical value of $\mathcal{E}=0.2$ and $C=-1$ (brown solid line); brown dashed line is the same as in (a) and  corresponds to $\mathcal{E}=0.2$ and $C=-2.6$.  } \label{fig:FIG2}
\end{figure}

The longitudinal component of the Lorentz force that determines the evolution of $\tilde{p}_x$ involves an oscillating factor $\cos\psi\cos\phi$. 
Near the resonance, $\omega_{\beta}\approx\langle \omega_D\rangle$, we can can use an approximation $\langle\cos\psi\cos\phi\rangle\approx \alpha\cos{ \langle{\theta}\rangle}$,  where  $\alpha\approx 0.348$, see Appendix and Ref.[]. After averaging out the betatron oscillations, the first equation in (\ref{eq:Eq_AA_2}) reduces to
\begin{eqnarray}
 \dot{\tilde{p}}_x =-4\alpha \mathcal{E}\cos{ \langle{\theta}\rangle}/{\tilde{p}_x}^{1/2}. \label{eq:EqM_AV}
\end{eqnarray} 
 Dividing Eq.~(\ref{eq:Eq_theta}) by  Eq.~(\ref{eq:EqM_AV}) we eliminate the dependence on time
\begin{eqnarray}
\frac{d \sin  \langle{\theta}\rangle}{d \tilde{p}_x}= -\frac{1}{4\alpha\mathcal{E} }\bigg({1-\frac{1}{\tilde{p}_x^{1/2}}} \bigg).
 \label{eq:F21}
\end{eqnarray}
 We then find that the trajectory in  phase space $(\tilde{p}_x, \theta)$ is given by: 
\begin{eqnarray}
 \sin  \langle{\theta}\rangle=\frac{1}{{4\alpha \mathcal{E}}}H(\tilde{p}_x)+ C,\quad H(\tilde{p}_x)\equiv {2\tilde{p}_x^{1/2}-\tilde{p}_x},
\label{eq:TRAJECT_EXP0}
\end{eqnarray}
where $C$ is  an integration constant determined by electron initial conditions. 

We can now determine the maximum longitudinal momentum that can be achieved by an electron that is initially at rest. The function $H$ increases with $\tilde{p}_x$ for $\tilde{p}_x \leq 1$ and then monotonically decreases for $\tilde{p}_x > 1$. Since $\sin \theta$ cannot exceed 1, then having $\sin \theta = 1$ at $\tilde{p}_x = 1$ allows the electron to reach the maximum possible momentum moving along a trajectory described by Eq. (18) in the $(\tilde{p}_x, \theta)$-phase space. Examples of such trajectories are shown in Fig.~\ref{fig:FIG2}(a). The momentum increases until $\sin \theta$ becomes equal to $-1$. This gives us the following condition for determining the maximum momentum $\tilde{p}_{\max}$:
\begin{eqnarray}
\frac{1}{4\alpha \mathcal{E}}H(\tilde{p}_x)|_{\tilde{p}_x=1}-\frac{1}{4\alpha \mathcal{E}}H(\tilde{p}_x)|_{\tilde{p}_x=\tilde{p}_{\max}}=2 
\label{eq:equationPmax}
\end{eqnarray}
which readily yields $\tilde{p}_{\max}= (1+\sqrt{8\alpha \mathcal{E}})^2$.

The trajectories in Fig.~\ref{fig:FIG2}(a) are plotted for three different values of $\mathcal{E} = 0.2$, $\mathcal{E} =0.36$, and $\mathcal{E} = 0.54$ and they illustrate an important topological change that takes place in the momentum space with the increase of this parameter. At $E = 0.2$, the electron momentum never reaches zero, whereas, at $E \geq 0.36$, the minimum electron momentum is zero. Using the quadratic equation (\ref{eq:equationPmax}) we can find the critical value $E_{cr}$ that corresponds to this transition:  $\mathcal{E}_{cr} =1/(8\alpha)=0.36$. Also it follows from this equation that for $\mathcal{E}\leq \mathcal{E}_{cr}$ the minimum momentum is $\tilde{p}_{\min}= (1-\sqrt{8\alpha \mathcal{E}})^2$. Therefore, we arrive to an important conclusion that initially cold electrons can achieve $\tilde{p}_{\max}$ only at $\mathcal{E} \geq \mathcal{E}_{cr}$. Electrons must be preheated at $\mathcal{E} <\mathcal{E}_{cr}$ in order to achieve $\tilde{p}_{\max}$ determined from Eq.~(\ref{eq:equationPmax}).

  
This observation raises a question of how an initially cold electron would move in the $(\tilde{p}_x, \theta)$-phase space at $\mathcal{E} < \mathcal{E}_{cr}$. Such an electron starts its motion at $\tilde{p}_x = 0$ with a given initial phase $\theta_{in}$ that determines the integration constant in Eq.~(\ref{eq:TRAJECT_EXP0}), $C = \sin \theta_{in}$. The highest value of the longitudinal momentum, $\tilde{p}_{x}=[1-(1-4(1-\sin\theta_{in})\alpha \mathcal{E})^{1/2}]^2$, is reached at  $\theta  = 1$. Therefore, for a given initial phase $\theta_{in}$, the longitudinal momentum of an initially cold electron would always oscillate in the interval $0<\tilde{p}_{x}<\tilde{p}_{sub}\equiv [1-(1-{\mathcal{E}/\mathcal{E}_{cr})^{1/2}}]^2<1$. An example of such a trajectory is shown in Fig.~2(b) as a left part of the  solid curve. It is worth pointing out that if we formally allow $\tilde{p}_x$ to increase in Eq.~(\ref{eq:TRAJECT_EXP0}), then eventually at $\tilde{p}_x > 1$ the right-hand side of this equation would again become smaller than unity. This would correspond to a solution also shown in Fig.~\ref{fig:FIG2}(b) with a right part of the solid curve. In this case, the electron longitudinal momentum oscillates while remaining greater than 1. 

The example shown in Fig.~\ref{fig:FIG2}(b), clearly illustrates that at $\mathcal{E} <\mathcal{E}_{cr}$ there appears a forbidden zone for an initially cold electron where formally $\sin \theta > 1$.  This zone cannot be crossed by initially cold electrons, see Fig.~\ref{fig:FIG2} (b), which perform only small amplitude oscillations along the left part of the phase trajectory which ends at the point $(\tilde{p}_x,\sin\theta)=(\tilde{p}_{sub},1)$. The presence of the forbidden zone leads to a threshold dependence of $\tilde{p}_{\max}$ on the parameter $\mathcal{E}$. Indeed, as we have already shown, we have

\begin{eqnarray}
  \begin{aligned}
{\max}\,{p}_x=m_ec\mathcal{I}_0\frac{\omega_L^2 }{2\omega_p^2}
 \big(1+\sqrt{8\alpha \mathcal{E}}\big)^2, \, \text{if}\ \mathcal{E}\ge \mathcal{E}_{cr}, \\
  {\max}\,{p}_x<m_ec\mathcal{I}_0\frac{\omega_L^2 }{2\omega_p^2}, \qquad\qquad\quad \, \text{if}\  \mathcal{E}< \mathcal{E}_{cr}.
	  \end{aligned}
	\label{eq:thresholdPmax}
\end{eqnarray}

\subsection{Results of numerical analysis}

In order to verify the predictions of Sec. \ref{SecA} obtained in 
 the paraxial approximation by averaging out betatron oscillations, 
 we have solved numerically the exact equations of motion (\ref{eq:EqM_N1}) - (\ref{eq:EqM_N3})  for initially cold electrons and a wide range of laser and plasma parameters. The numerical solutions confirm that indeed the phase trajectories obtained from the exact equations of motion are close to the averaged trajectories even at a small number of betatron oscillations, see Fig~3(a) and 3(b). The numerical solutions also confirm that all electron trajectories in the $(p_x, \theta)$-phase space can be divided into two distinct groups described in the previous section depending on the value of $\mathcal{E}$. For the first group of trajectories, the momentum $p_x$ oscillates while remaining below $p_*$, while, for the second group of trajectories, the momentum $p_x$ can increase dramatically along the phase trajectories and becomes well above $p_*$. The numerically calculated critical value of $\cal{E}$ that separates the two groups is $\mathcal{E}^{(num)} = 0.54$. This value is somewhat above than what was predicted by our theory. One of the reasons for this is that the  paraxial approximation is only applicable at ultra-relativistic energies and that is why an exact solution is needed to quantitatively predict the threshold value of $\mathcal{E}$. Figure~4(a) illustrates a significant change in the energy gain at values of $\mathcal{E}$ close to $\mathcal{E}_{cr}$.


\begin{figure}[t]
 \centering
\includegraphics[height=0.12\textheight,width=0.85\columnwidth]{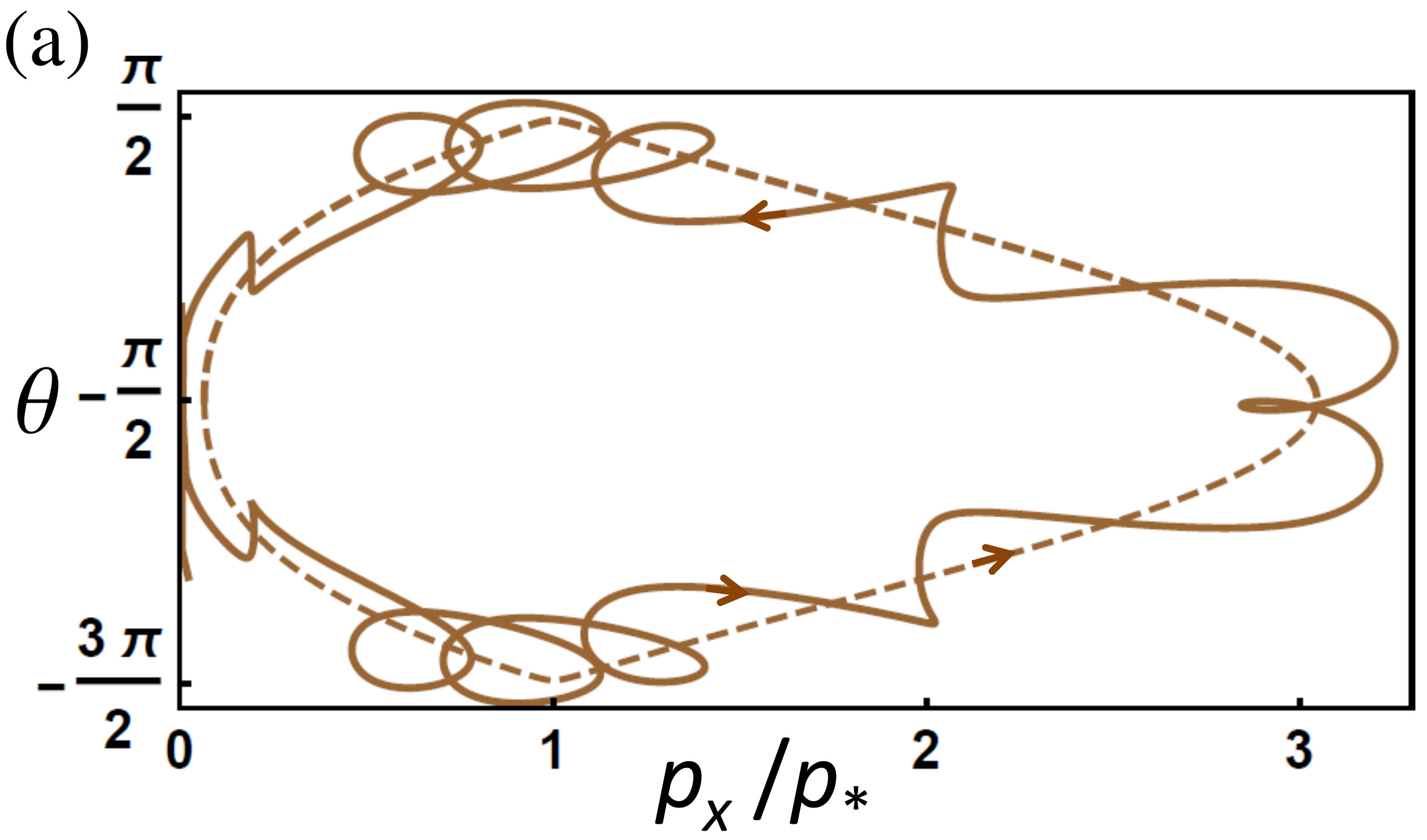}\\
\includegraphics[height=0.13\textheight,width=0.85\columnwidth]{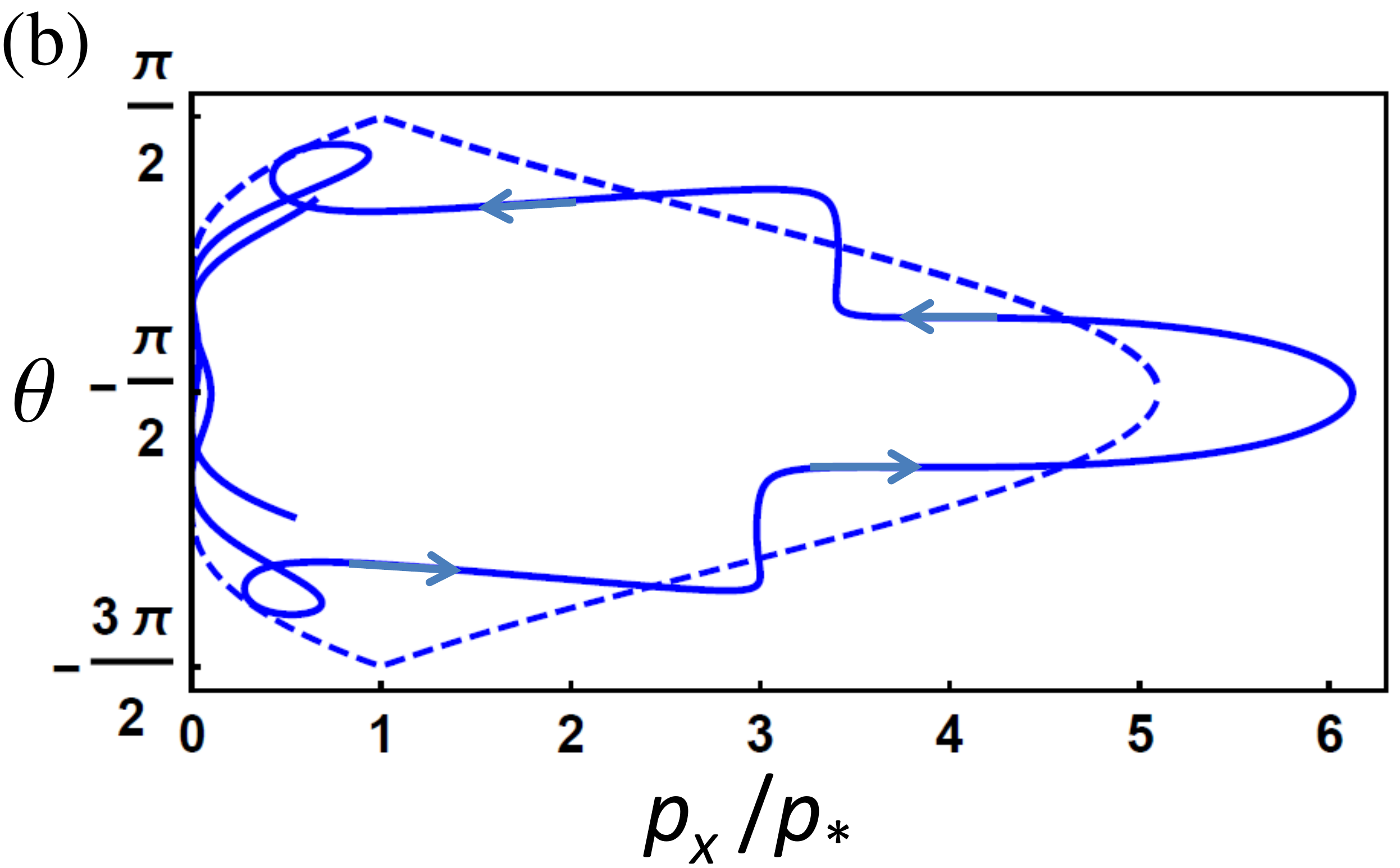}
\caption{ Comparison of  phase trajectories obtained from exact  (solid) and  averaged   (dashed) equations of motion at 
different $\mathcal{E}$:
(a)  $\mathcal{E}=0.2$ and (b)  $\mathcal{E}=0.57$.
The corresponding $p_x/p_*$ profiles versus time $t$ are  shown in Fig.~1 (brown curve) and Fig.~\ref{fig:FIG4}(a)  (blue curve).
  }  \label{fig:FIG3_}
\end{figure}

We have repeated the numerical analysis for different initial positions of the electron. As previously stated, by varying the initial longitudinal location $x$, we effectively vary the initial phase $\phi$ of the wave. In this case the parameter $\mathcal{E}$ remains the same. However, the change in the initial phase $\phi$ can have a profound impact on the electron dynamics in the ion channel. Figure~4(b) shows the maximum relativistic $\gamma$-factor attained by an electron as a function of $\cal{E}$ for two different initial phases. In the first case, the electron begins its motion at $\phi = 0$, so that initially $E = E_0$ and $a = 0$. In the second case, the electron begins its motion at $\phi = \pi/2$, so that initially $E = 0$ and $a = a_0$. We find that the position of the threshold changes significantly with the initial phase. 
 For comparison, we have also plotted in this figure the maximum energy gain  obtained from Eq.~(\ref{eq:equationPmax}) which is transformed to the following convenient form:
  \begin{eqnarray}
\frac{\gamma_{\max}}{\gamma_*}=\frac{1}{\mathcal{E}^2}(1+\sqrt{8\alpha\mathcal{E}})^2.
\label{eq:gamma_max}
\end{eqnarray}
 where $\gamma_*\equiv a_0^2/2\mathcal{I}_0$ is  the upper limit for the energy gain from the laser in a vacuum  at large $a_0$. This formula gives an  amplification coefficient of the electron energy introduced by the ion channel.

In order to understand the underlying cause for the threshold change, it is worth considering a simple case of an electron in a vacuum. The momentum components in this case are $p_y = a-a_{in}$ and $p_x= p_y^2/2$, where $a_{in}$ is the initial amplitude of the vector potential  $a$ determined by the initial phase $\phi$. Therefore, the maximum amplitude of the $p_x$-oscillations induced by the wave increases from $a_0^2/2$ at $a_{in} = 0$ to four times this value at $a_{in} = -a_0$. The change in phase can then be viewed as electron preheating that was discussed in the previous section. In other words, having stronger oscillations of $p_x$ allows the electron to access those averaged trajectories for which $p_{\min} > 0$. 
 Different initial wave phases    can be realized, for example, when electrons are injected in the channel through ionization of the doping gas.

Additional numerical calculations show that the estimate given by Eq.~(\ref{eq:gamma_max}) provides an accurate upper limit for the electron energy gain regardless of the initial wave phase. It remains accurate even when the number of oscillation during the energy gain is not very large. We find that the number of transverse oscillation during the energy gain can be relatively well approximated by $N_c\sim 1.7{\mathcal{E}}^{-1/2}$. The corresponding longitudinal distance travelled by the electron is then given by $D\sim  c(2\pi/\omega_*)  N_c \sim 1.7{\mathcal{E}}^{-1/2} \mathcal{I}_0(\omega_L/\omega_p)^2\lambda$, where $\lambda=c \pi/\omega_L$.

\begin{figure}[t]
 \centering
\includegraphics[height=0.12\textheight,width=0.9\columnwidth]{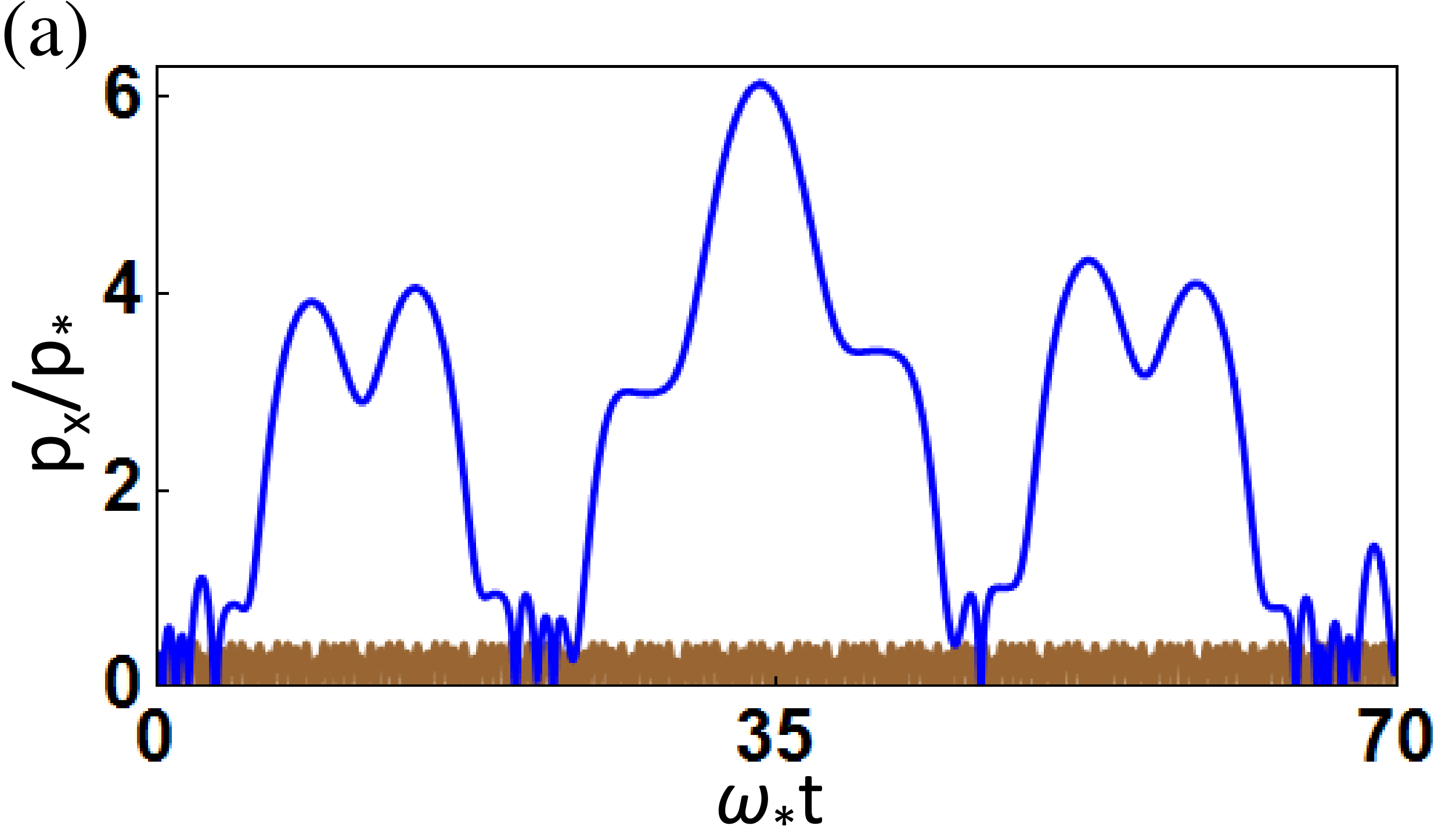}
\\
\,\,\, 
\includegraphics[height=0.120\textheight,width=0.90\columnwidth]{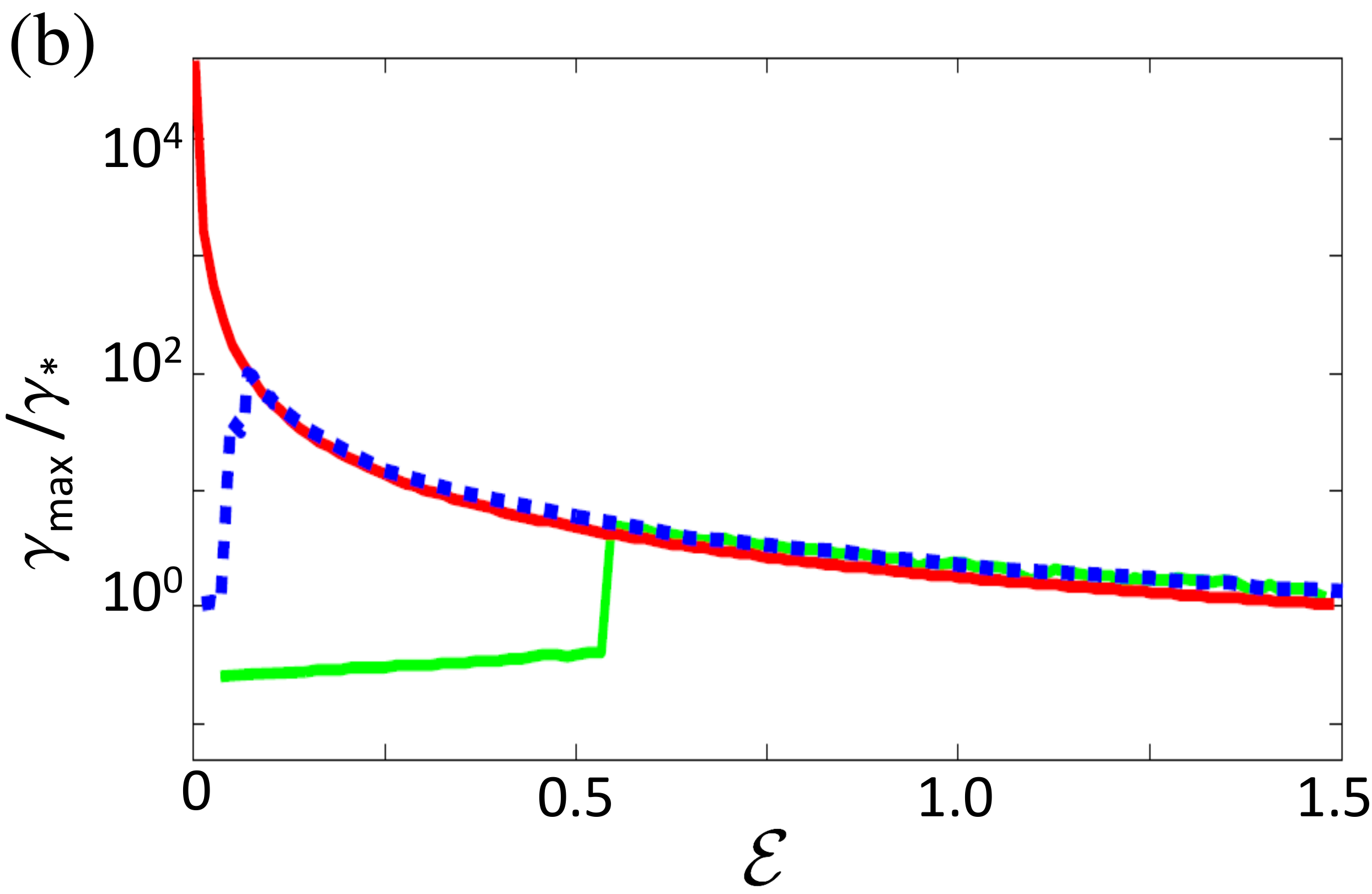}
\caption{(a) Change in electron dynamics at  near-critical values of  $\mathcal{E}$:  $\mathcal{E}=0.57>\mathcal{E}_{cr}=0.54$ with $\omega_p/\omega = 0.114$ and  $a_0 =5$  (blue);   $\mathcal{E}=0.513<\mathcal{E}_{cr}$ with $\omega_p/\omega = 0.114$ and  $a_0 =4.5$ (brown).
(b) Maximum energy gain $\gamma_{\max}/\gamma_*$  as a function of $\mathcal{E}$ at  $\phi|_{t=0}=0$ (green) and $\phi|_{t=0}=\pi/2$ (dashed blue). The red curve is the estimate for the upper limit of $\gamma_{\max}$ given by Eq.~(\ref{eq:gamma_max}). } \label{fig:FIG4}
\end{figure}

 
We have so far considered the case in which electrons start their motion on the axis of the channel, so that the resulting electron trajectories driven by the laser field are flat. An initial displacement out of the plane of the driven oscillations leads to a three-dimensional electron trajectory []. Solving Eqs.~(\ref{eq:EqM_N1}) - (\ref{eq:EqM_N3}) numerically in this regime, we found that the maximum electron energy gain is close to that predicted for the planar trajectories. During the energy gain, the electron trajectory in the cross-section of the channel resembles an ellipse considerably elongated along the laser electric field. 
\section{Scalings in the super-luminal case}
In the luminal case $v_{ph}=c$ considered in Sec.~III, the longitudinal  
velocity of ultrarelativistic electrons is  separated on average from the speed of light
by $c-\langle v_x\rangle=\langle\omega_D/\omega_L\rangle c\sim  
c/p_*\sim (\omega_p/\omega_L)^2 c$. When electrons are accelerated by  
the super-luminal wave, this difference can be comparable  to 
$v_{ph}-c\sim \chi(\omega_p/\omega_L)^2 c$, and therefore one can expect a  
strong influence of  super-luminosity on the resonant interaction  
between electrons and the wave~[].

In what follows, we use an approach developed in the previous section  in order to find how the phase shift between a super-luminal wave and transverse electron oscillations in the channel evolves with time. In the super-luminal case ($\chi>0$), the  energy of the transverse  oscillations grows with the longitudinal electron momentum according to Eq.~(\ref{eq:Energy}). The amplitude of these oscillations also grows with $\tilde{p}_x$ as ${\tilde{y}}=(1+\chi\tilde{p}_x)^{1/2}\sin\psi$. Under an assumption that ${\tilde{p}_x}$ increases over many betatron oscillations, we find that the transverse velocity in the super-luminal case is determined by $\dot{\tilde{y}} \approx  (1+\chi\tilde{p}_x)^{1/2}\cos\psi \dot{{\psi}}$, where $\dot{{\psi}}=1/{\tilde{p}_x}^{1/2}$. After averaging over betatron oscillations, Eq.~(\ref{eq:Eq_AA_3}) for the wave phase takes the form: 
$\langle{\dot{\phi}}\rangle=-2\langle{\dot{\tilde{y}}^2}\rangle-2\chi=-1/\tilde{p}_x-3\chi$. Combining the expressions for $\langle{\dot{\phi}}\rangle$ and $\dot{{\psi}}$ we then find that the averaged phase shift evolves with time as 
\begin{eqnarray}
\langle\dot{\theta}\rangle=\frac{\omega_{\beta}-\langle{\omega_D}\rangle}{\omega_*}=\frac{1}{\tilde{p}_x^{1/2}}-\frac{1}{\tilde{p}_x}-3\chi.
  \label{eq:phi_SUPER}
\end{eqnarray}
\begin{figure}[t]
  \centering
\quad
\includegraphics[height=0.1200\textheight,width=0.87\columnwidth]{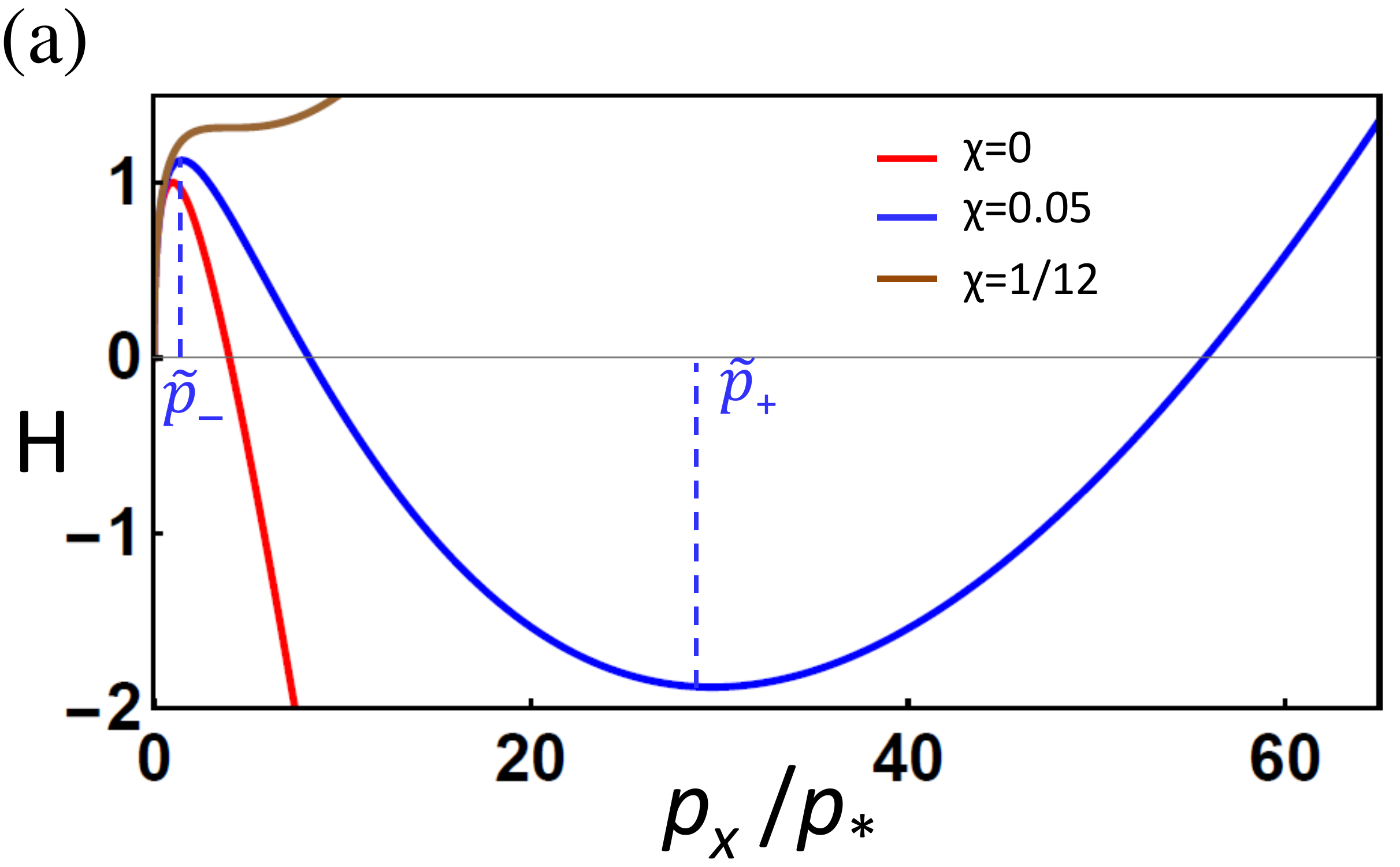}\\
\includegraphics[height=0.1300\textheight,width=0.90\columnwidth]{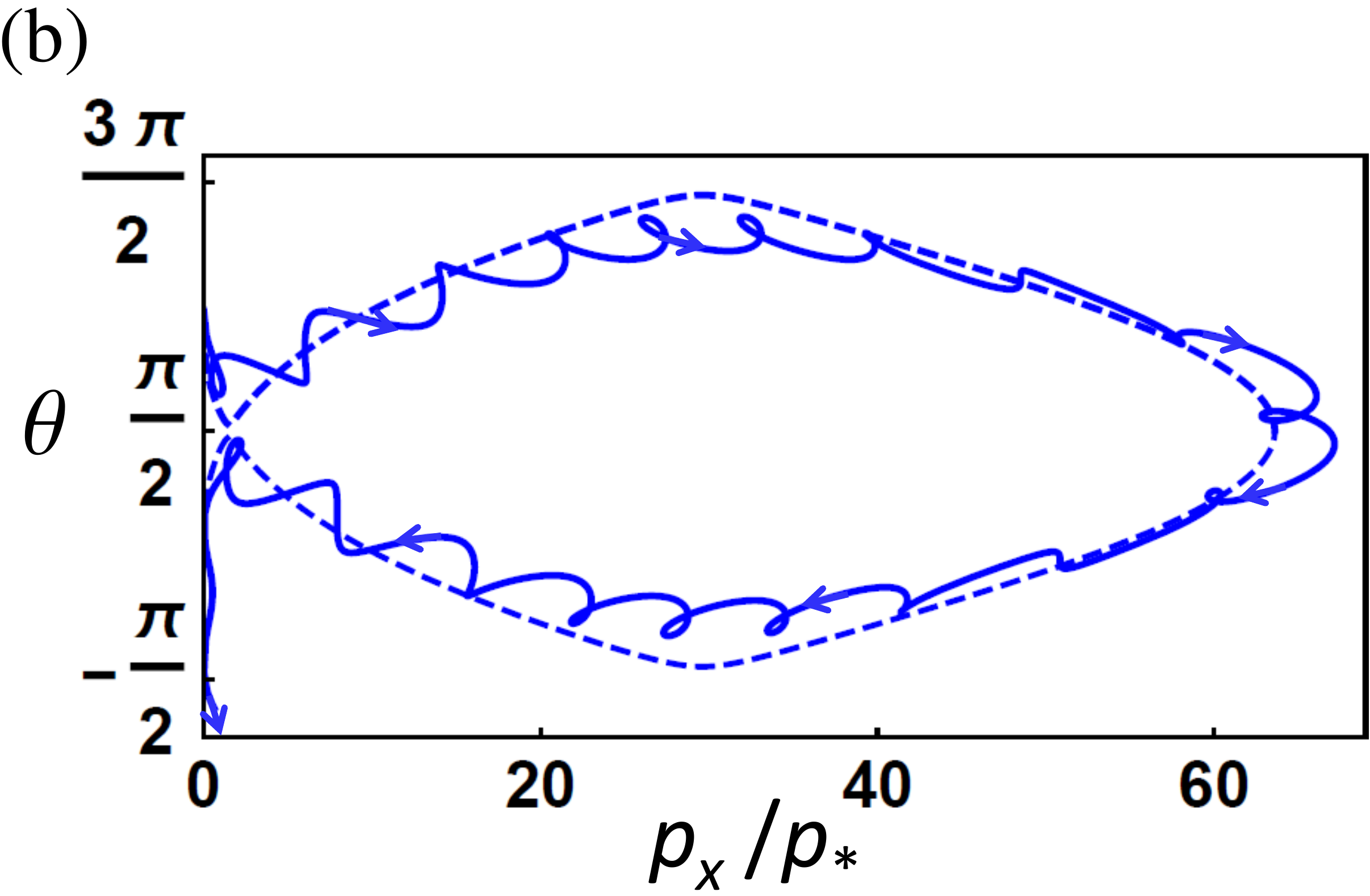}
\caption{(a) Dependence $H(\chi,{\chi\tilde{p}_x})$ on  
$\tilde{p}_x=p_x/p_*$ at different values of $\chi$. When  
$\chi\rightarrow 0$, $\tilde{p}_{+}\rightarrow \infty$ and  
$\tilde{p}_{-}\rightarrow 1$ ($\tilde{p}_{x}=1$ is the position of  
maximum on red curve); when $\chi\rightarrow 1/12$,  
$\tilde{p}_{+}\rightarrow 4$ and  $\tilde{p}_{-}\rightarrow 4$  
($\tilde{p}_{x}=4$ is the position of inflection point on the brown  
curve). (b) Exact (solid) and averaged (dashed) phase trajectories for  
initially cold electrons: $\mathcal{E}=1.092$ and $\chi=0.05$  
($a_0=6$, $\omega_p/\omega_L=0.182$, and $v_{ph}=1.00331c$). Maximum  
energy $\gamma_{\max}\sim 67 p_*/m_ec\approx 1000$. } \label{fig:FIG5}
\end{figure}

Close to the resonance, $\omega_{\beta}\approx \langle{\omega_D}\rangle$,  
the equation for the longitudinal momentum [see Eq.~(11)] takes the following form:
\begin{eqnarray}
  \dot{\tilde{p}}_x=-4\alpha \mathcal{E}\cos{  
\langle{\theta}\rangle}[(1+\chi \tilde{p}_x)/\tilde{p}_x]^{1/2}.
  \label{eq:px_SUPER}
\end{eqnarray}
Eliminating the explicit dependence on time by dividing Eq.~(\ref{eq:phi_SUPER}) by  
Eq.~(\ref{eq:px_SUPER}) we obtain
\begin{eqnarray} \label{eq:EQ_PH_TR_SUPER}
\frac{d }{d \tilde{p}_x}\sin \langle\theta\rangle=  
-\frac{1-1/\tilde{p}_x^{1/2}-3\chi\tilde{p}_x^{1/2}}{4\alpha  
\mathcal{E} \sqrt{1+\chi\tilde{p}_x}}.
\end{eqnarray}
Electron trajectories in the $(\tilde{p}_x,\theta)$-phase-space are then specified by
\begin{eqnarray}
\sin  \langle{\theta}\rangle=\frac{1}{4\alpha  
\mathcal{E}}H(\chi,{\chi\tilde{p}_x})+C,\,\,\,\label{eq:PH_TR_SUPER}\\
H(\chi,z)\equiv  
\frac{3\sqrt{z}\mu(z)-\ln[\sqrt{z}+\mu(z)]}{\sqrt{\chi}}-\frac{2[\mu(z)-1]}{\chi},\,\,\,  
\label{eq:H_SUPER}
\end{eqnarray}
where $\mu(z)\equiv \sqrt{z+1}$.

In contrast to the luminal case, the resonance condition $\langle{\omega}_D\rangle = \omega_{\beta}$ can be satisfied in the super-luminal case for two values of the longitudinal momentum, $\tilde{p}_{\pm}=(1/6\chi)^2(1\pm\sqrt{1-12\chi})^2$ when  the parameter $\chi$ that characterizes the super-luminosity lies in the range $0<\chi\leq 1/12$. As illustrated in Fig.~\ref{fig:FIG5} (a), the function $H(\chi,{\chi\tilde{p}_x})$ has a maximum at  
$\tilde{p}_x=\tilde{p}_{-}$ and a minimum at  $\tilde{p}_x=\tilde{p}_{+}$. It is important to note that at small $\chi$ the second resonance occurs at much larger values of the longitudinal momentum, $\tilde{p}_{+}\propto \chi^{-2}\gg  1$, than in the luminal case. As a result, the super-luminosity enables a considerable energy increase compared to the luminal case. Figure~\ref{fig:FIG5}~(b), where maximum ${p}_x$ exceeds $p_*$ by a factor more than 60, illustrates this aspect.

\begin{figure}[t]
  \centering
\quad\quad\includegraphics[height=0.1600\textheight,width=0.95\columnwidth]{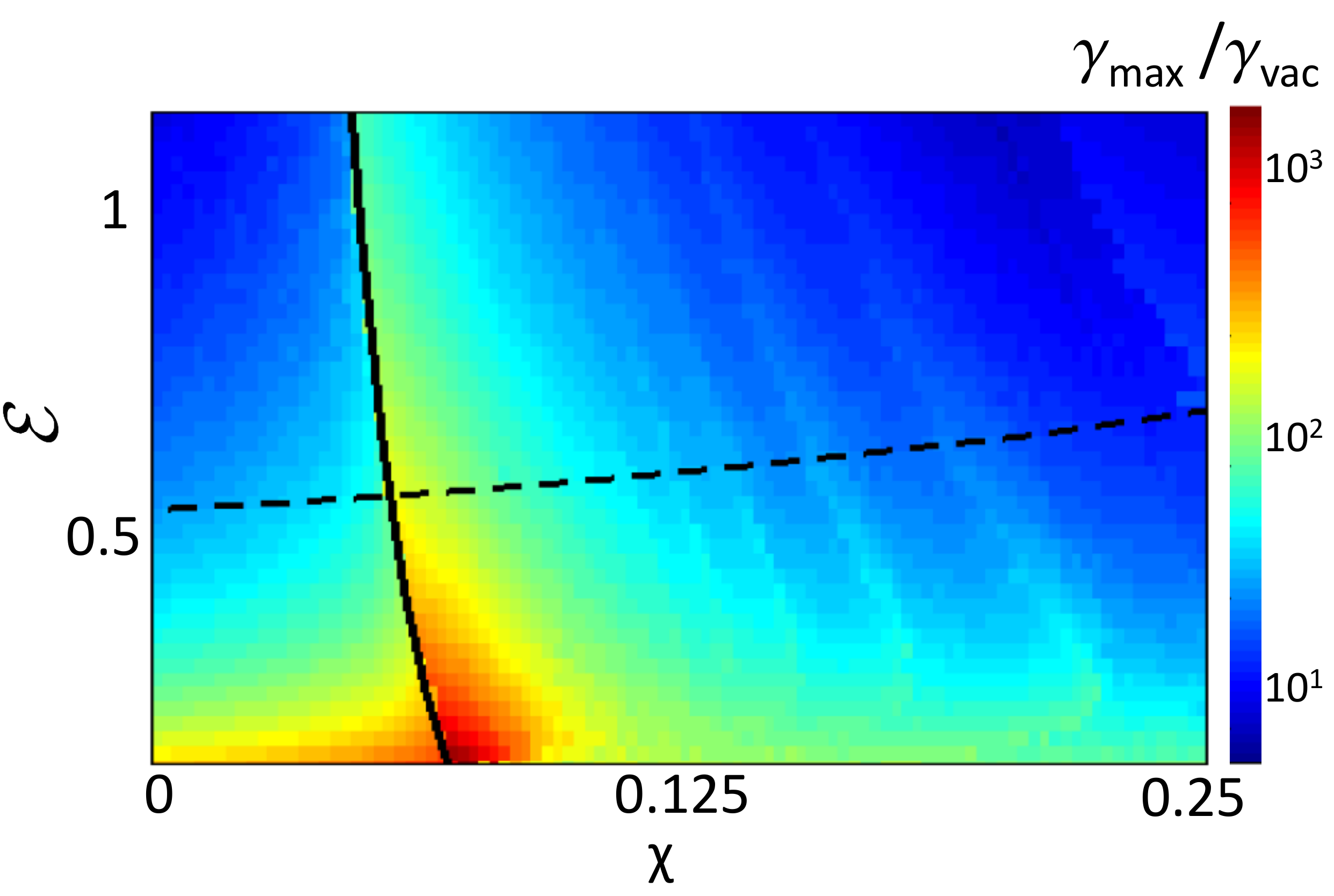}
\caption{   Color-coded  upper limit for maximum energy  as a  
function of $\mathcal{E}$ and $\chi$. The energy area below dashed  
line is reachable only for initially preheated electrons (that is,  at  
$\phi|_{t=0}=\pi/2$). The area right of the black solid line  
corresponds to the range of laser-plasma parameters at which  
$[H(\chi,\chi\tilde{p}_{-})-H(\chi,\chi\tilde{p}_{+})]/4\alpha  
\mathcal{E}\leq 2.2$.} \label{fig:Universal}
\end{figure}

As evident from Eqs.~(\ref{eq:PH_TR_SUPER}) and (\ref{eq:H_SUPER}), the maximum longitudinal momentum now depends on two parameters, $\tilde{p}_{\max}=\tilde{p}_{\max}(\chi, \alpha  
\mathcal{E})$. An upper limit for the electron energy gain is then given by  
\begin{eqnarray}
\frac{\gamma_{\max}}{\gamma_*}=\frac{1}{\mathcal{E}^2}\tilde{p}_{\max}(\chi,\alpha  
\mathcal{E}),\label{eq:P_MAX_SUPER}
\end{eqnarray}
where $\gamma_*\equiv a_0^2/2\mathcal{I}_0$. We have evaluated $\gamma_{\max}/\gamma_*
$ numerically by performing a wide parameter scan, with the result of the scan shown in Fig.~\ref{fig:Universal}. The dashed lines marks an energy gain threshold, dividing the parameter-space ($\chi,\mathcal{E}$) into two distinct areas. For parameters above the dashed line, electrons gain the same maximum energy with and without initial preheating (that is, at  $\phi|_{t=0}=p/2$ and  at $\phi|_{t=0}=0$). For parameters below the  dashed line, only initially preheated  electrons ($\phi|_{t=0}=p/2$) gain the energy given by Eq.~(\ref{eq:P_MAX_SUPER}), whereas initially cold electrons ($\phi|_{t=0}=0$) gain considerably less energy. This threshold matches at $\chi = 0$ the threshold discussed in Sec. IIIA for the luminal case.

The vertical solid curve in Fig.~\ref{fig:Universal} marks a threshold introduced by the super-luminosity of the wave. As shown in Fig.~\ref{fig:FIG5} (a), the function $H(\chi,{\chi\tilde{p}_x})$ that determines the time evolution of the phase shift $\theta$ can have a minimum in addition to a maximum that is also present in the luminal case. This minimum is accessible to an electron during its acceleration as $\tilde{p}_x$ continues to increase, but only if the difference  between the maximum and the minimum is not too large,
\begin{eqnarray}
[H(\chi,\chi\tilde{p}_{-})-H(\chi,\chi\tilde{p}_{+})]/4\alpha  
\mathcal{E}\leq 2.
\label{eq:2EXTREMUM}
\end{eqnarray}
A detailed numerical analysis shows that the constant on the right-hand side of this inequality is 2.2 rather than~2. The condition given by Eq.~(\ref{eq:2EXTREMUM}) [with 2.2 on right-hand side] is satisfied for the parameters in the area to the right of the solid line in Fig.~\ref{fig:Universal}.  Immediately to the right of the solid line, we have $\gamma_{\max}/\gamma_{vac} \sim 2\mathcal{E}^{-2}\tilde{p}_{+}$. The maximum energy drops by approximately a factor of two to the left of the threshold.

The analysis presented here demonstrates that super-luminosity can have a positive impact on electron energy gain during electron acceleration in an ion channel. Effective energy gain requires for the betatron frequency $\omega_{\beta}$ to remain close to the Doppler shifted frequency $\langle\omega_D\rangle$ with the increase of the electron's longitudinal momentum. In the luminal case, we have $\omega_{\beta} \propto 1/\tilde{p}_x^{1/2}$ and $\langle\omega_D\rangle \propto 1/\tilde{p}$, that is,  the Doppler shifted frequency decreases with the longitudinal momentum faster than the betatron frequency. This eventually leads to a breakdown of the resonance condition and stems the longitudinal momentum gain. The superluminal correction does not change the betatron frequency,  but  it does increases  the Doppler shifted frequency:   $\langle\omega_D\rangle\propto (1/\tilde{p}_x+3\chi)$. As a result, accelerating electrons can stay in resonance with the laser wave at higher values of the longitudinal momentum and thus gain significantly more energy than in the luminal case.

 We have found that  three dimensional effects in the superluminal case are similar to those in the luminal case: the particle  pushed out of the plane formed by the laser polarization and the channel axis moves along a three dimensional trajectory and its maximum  energy is the same as during planar motion. We also have observed the situations when  large deviations from the plane  can sharply decrease the particle energy gain. 



\section{Spinoffs  of universal scaling theory  }
The  developed approach is quite general and therefore it can be  
applied to examine several seemingly different regimes that we discuss  
in this section.
\subsection{Acceleration of electrons through the resonance with the  
third harmonic of betatron oscillations}
In the previous section, we considered electron acceleration and the resulting energy gain when the Doppler shifted frequency of the laser wave is close to the average frequency of betatron oscillations. However, efficient energy gain is also possible through a resonance of higher harmonics, with $|\langle \omega_D\rangle|\approx l\omega_{\beta}$, where $l$ is the number of the betatron harmonic.

In order to describe such a regime, one needs to  introduce only several changes. Equation~(\ref{eq:phi_SUPER}) for  the averaged phase shift, $ \langle{\theta}^{(l)}\rangle=l\psi+\langle\phi\rangle$, now takes the  
following form

\begin{eqnarray} 
\langle\dot{\theta}^{(l)}\rangle=\frac{l\omega_{\beta}-\langle{\omega_D}\rangle}{\omega_*}=\frac{l}{\tilde{p}_x^{1/2}}-\frac{1}{\tilde{p}_x}-3\chi.
  \label{eq:phi_SUPER_l}
\end{eqnarray}
After averaging the oscillating factor $\dot{\tilde{y}}\cos\phi$ in  
Eq.~(7), we find that $\langle\cos\psi\cos\phi\rangle\approx \alpha_l\cos \langle{\theta}\rangle$ near an $l^{\mbox{th}}$ harmonic resonance. The numerical constant $\alpha_l$ is non vanishing only for odd values of $l$ [], with  $\alpha_1 = 0.348101$, $\alpha_3 =-0.162924$, and $\alpha_5= 0.114729$. Therefore, the longitudinal momentum $\tilde{p}_x$ now satisfies Eq.~(\ref{eq:px_SUPER}), where $\alpha$ must be replaced with $\alpha_l$.

After straightforward transformations, we obtain an equation for the electron phase trajectory:
\begin{eqnarray}
\sin  \langle{\theta}^{(l)}\rangle=\frac{1}{4\alpha_l l  
\mathcal{E}^{(l)}}H[{\chi}^{(l)}/l^{2},{({\chi}^{(l)}/l^{2})(\tilde{p}_x}l^2)]+C,
\label{eq:TR_l_}
\end{eqnarray}
where $H$ is defined by Eq.~(\ref{eq:H_SUPER}). By comparing Eq. (\ref{eq:TR_l_}) to Eq.~(\ref{eq:PH_TR_SUPER}) that was derived for the main resonance, we can immediately conclude that an $l^{\mbox{th}}$ harmonic resonance is equivalent to the main resonance with
$\mathcal{E}=(|\alpha_l|/\alpha)l\mathcal{E}^{(l)}$ and $\chi=\chi^{(l)}/l^2$.
Therefore, the maximum energy for an $l^{\mbox{th}}$ harmonic resonance is given by
\begin{eqnarray}
\gamma_{\max}^{(l)}\approx \max\,  
\frac{p_x^{(l)}}{m_ec}=\frac{1}{l^{2}} \frac{p_{*}^{(l)}}{m_ec}\cdot  
p_{\max}({\chi},\mathcal{E}).
\label{eq:gamma_l}
\end{eqnarray}
\begin{figure}[t]
 \centering
	  \includegraphics[height=0.1200\textheight,width=0.85\columnwidth]{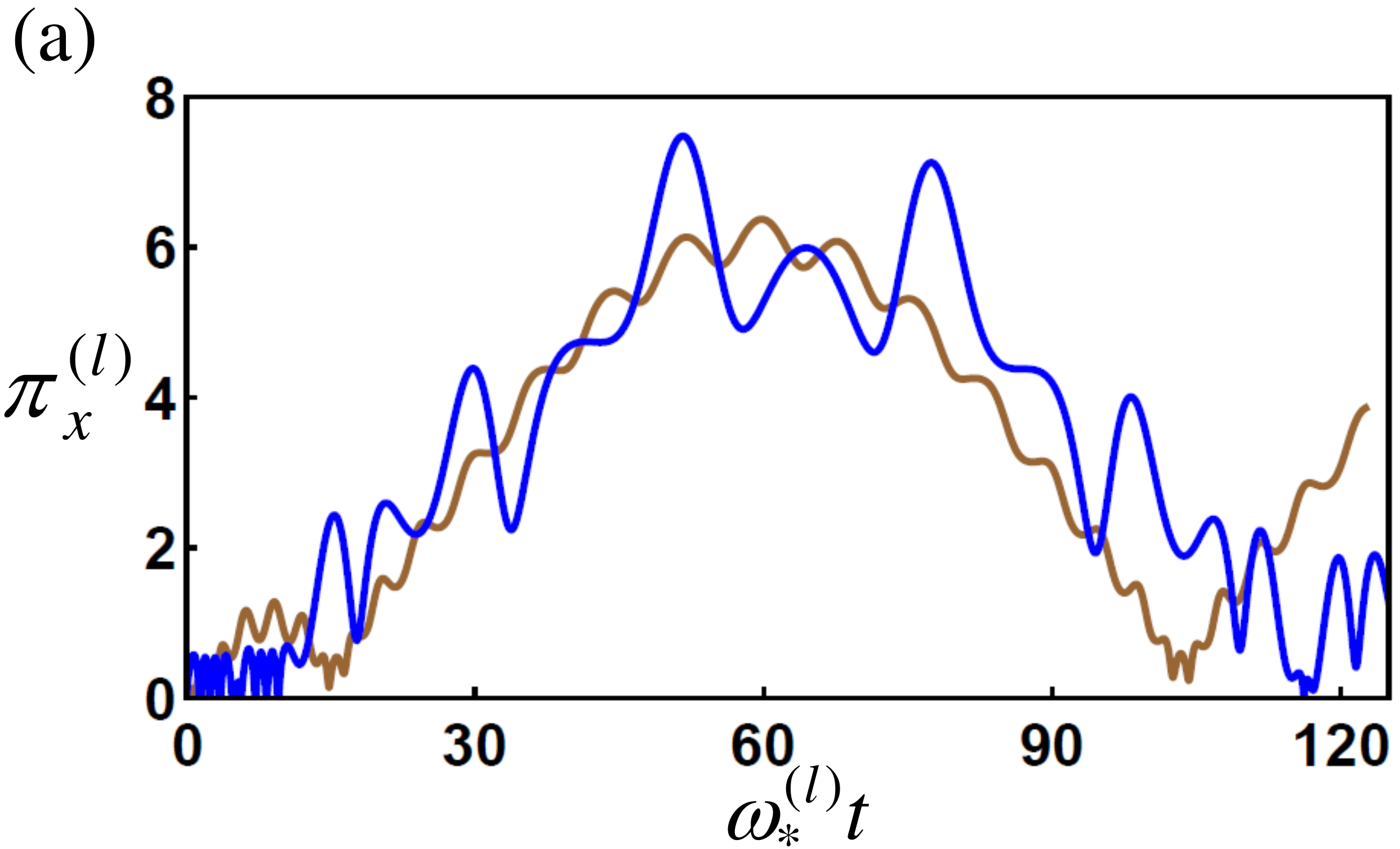}\\
		\includegraphics[height=0.1300\textheight,width=0.85\columnwidth]{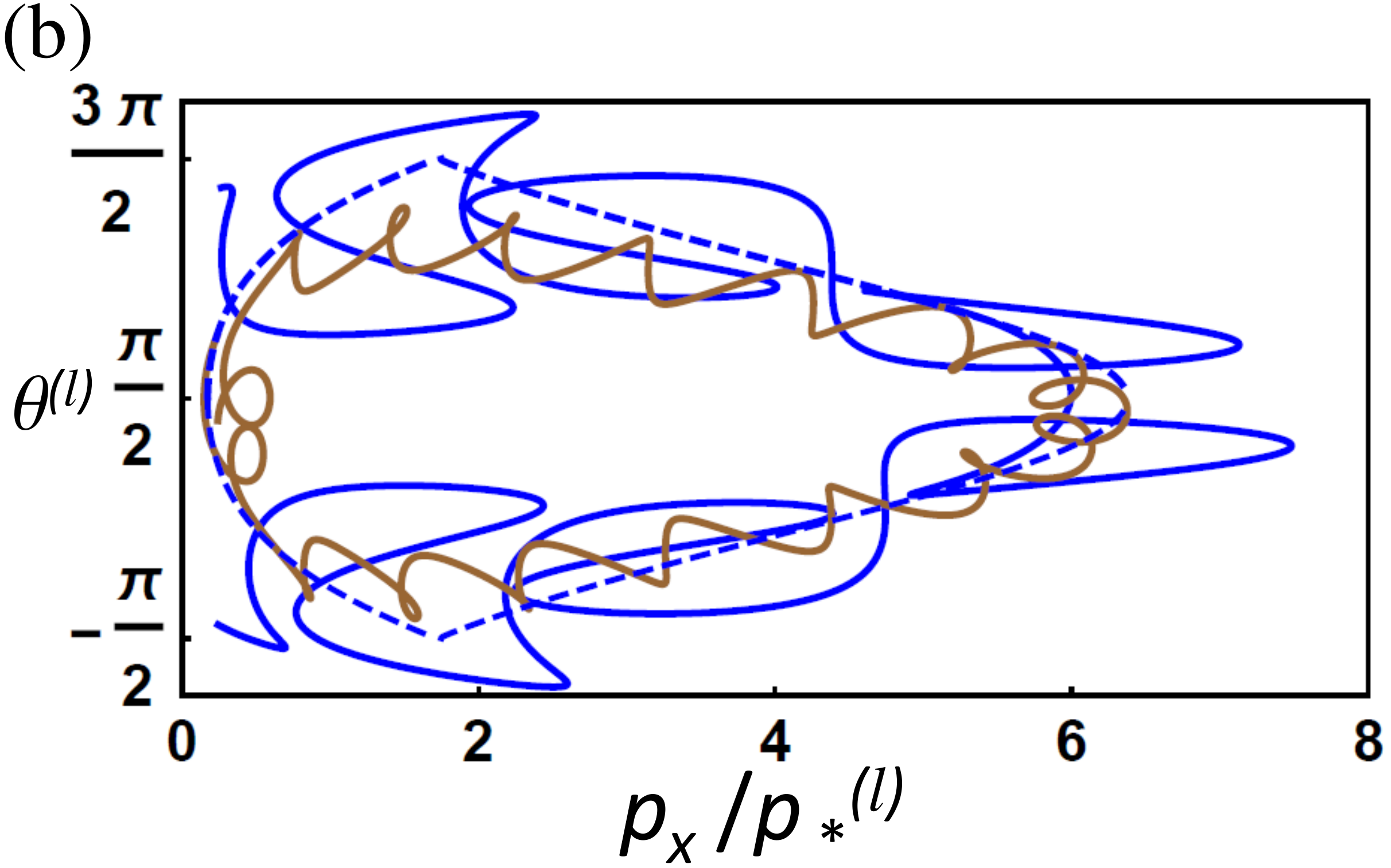}
			\caption{ (a) Dependence $\pi_x^{(l)}\equiv p_x/({p}_*^{(l)}/l^2)$  on time $\omega_*^{(l)}t$ for  $l=1$  (brown) and $l=3$ (blue) resonances at $(\chi^{(1)},\mathcal{E}^{(1)})=(0.06,0.18)$ and $(\chi^{(3)},\mathcal{E}^{(3)})=(0.54,0.128)$; $\omega_*^{(1)}\approx 0.0195\omega_L$, ${p}_*^{(1)}\approx 178 m_ec$ and $\omega_*^{(3)}\approx 0.088\omega_L$, ${p}_*^{(3)}/3^2\approx 39 m_ec$. (b) Phase trajectories for  the $l=1$  (brown) and   $l=3$ resonances (blue) at the same  $(\chi^{(1)},\mathcal{E}^{(1)})$ and $(\chi^{(3)},\mathcal{E}^{(3)})$ as in (a). Dashed blue line corresponds to averaged phase trajectory. In all cases, electrons are initially placed at rest on the channel cases.   \label{fig:traject_3res}}
\end{figure}

To illustrate the similarity between the main and the 3-harmonic resonances, we   
take $(\chi,\mathcal{E})\equiv (\chi^{(1)},\mathcal{E}^{(1)})=(0.06,0.18)$ for the main resonance, which corresponds to $p_{\max}({\chi},\mathcal{E})\approx 6.6$ according to the parameter scan shown in Fig.~\ref{fig:Universal}.  The matching parameters for  the third-harmonic resonance are $(\chi^{(3)},\mathcal{E}^{(3)})=(0.06 \cdot  3^2,0.18 \alpha/3|\alpha_3|)\approx(0.54,0.128)$. For these parameters, $p_{\max}$ is the same as for the main resonance. Assuming that in both cases  the electrons are initially placed off the channel axis at $y|_{t=0}=2\sqrt{6}c/\omega_p^{(l)}$ (that is, with $\mathcal{I}_0=7$) and accelerated by a wave with $a_0=9$, we find that $\omega_p^{(1)}/\omega_L=0.3695$,  
${v}_{ph}^{(1)}/c=1.00235$, and $p_*^{(1)}=178$ for the main resonance and $\omega_p^{(3)}/\omega_L=0.2608$,  ${v}_{ph}^{(3)}/c=1.0107$, and $p_*^{(3)}/3^2=39$ for the third harmonic resonance.

Although the overall normalized momentum gain is similar for the main and the third-harmonics resonance, see Fig.~\ref{fig:traject_3res}(a), $p_x(t)$ has a typical zig-zag signature in the case of the third harmonic resonance,  whereas $p_x(t)$ has smaller and smoother oscillations in the case of the main resonance. Electron phase-space trajectories shown in Fig.~\ref{fig:traject_3res}(b) for these two regimes also have distinctive signatures: the trajectory for the third harmonic resonance has large loop-like oscillations as compared to the trajectory for the main resonance.

In general, figure~\ref{fig:Universal} provides universal scalings for the electron energy gain via high harmonic resonances. However, the realization of these regimes depends on initial conditions (such as pre-acceleration) for given laser and plasma parameters.   
\subsection{Acceleration of electrons by the circularly polarized laser wave}
\begin{figure}[t]
  \centering
\includegraphics[height=0.17\textheight,width=0.85\columnwidth]{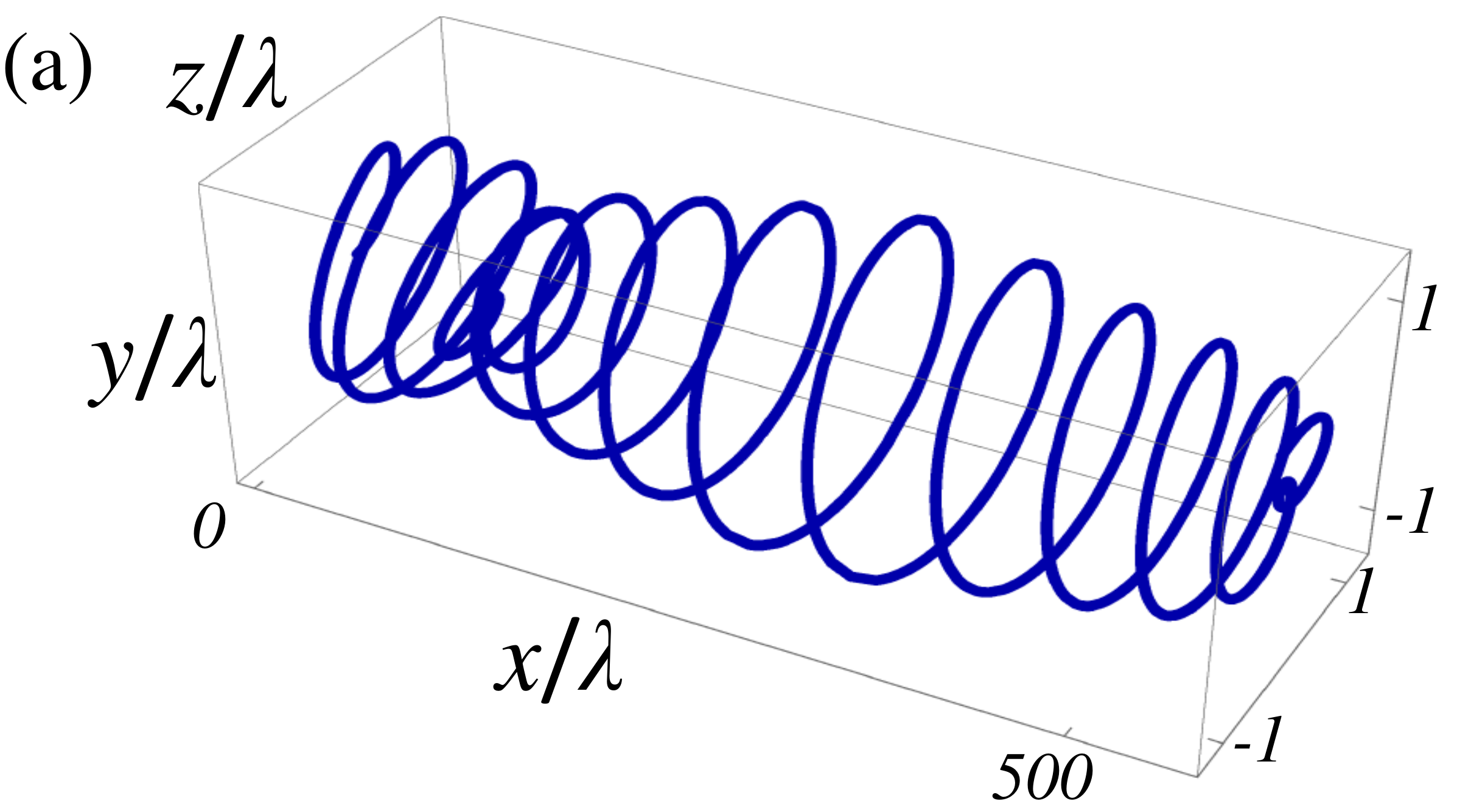}\\
\includegraphics[height=0.13\textheight,width=0.85\columnwidth]{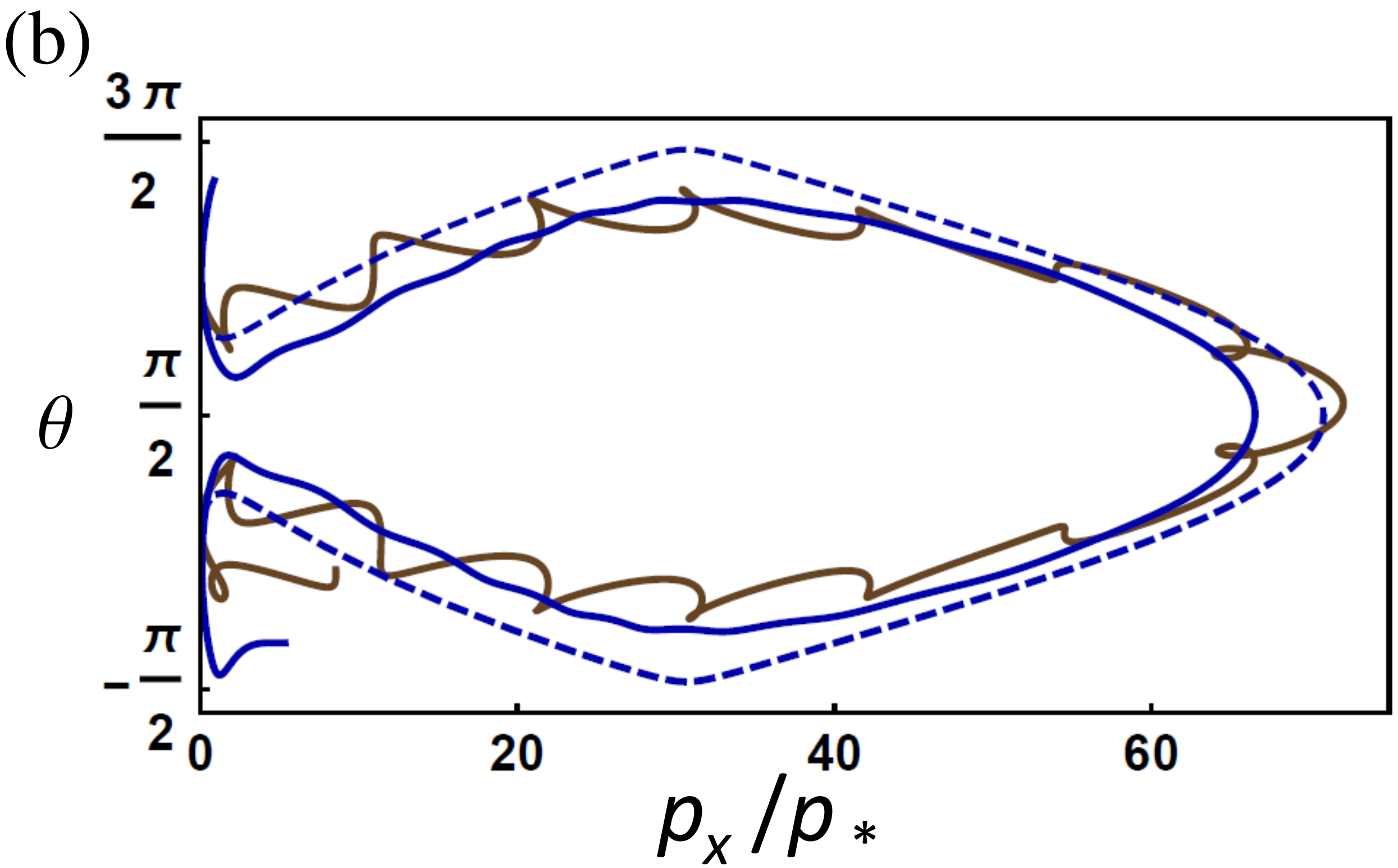}
\caption{(a) Trajectory of the relativistic electron  accelerated  
by the circularly polarized laser wave: $a_{0c}=1.5\sqrt{2}$,  
$\omega_p/\omega_L=0.333$, $v_{ph}=0.0111$, and $\gamma_{\max}\approx  
300$. (b) Phase trajectories obtained from exact equation of motion  
for electrons accelerated by circularly (solid blue) and linearly  
(brown) polarized laser wave with matching wave amplitude $a_c=\alpha  
\sqrt{2}a_0$. Dashed blue line corresponds to averaged phase  
trajectory. In all cases, electrons are initially placed at rest on  
the channel axis.    \label{fig:circular}}
\end{figure}

The other application of the universal scaling theory is the electron motion in a circular polarized (CP) laser wave. To be specific, we set ${\bf{E}}^{(L)} = E_{0c}({\bf{e}}_y\cos\phi+{\bf{e}}_z\sin\phi)$ and ${\bf{B}}^{(L)} = E_{0c}(c/v_{ph})({-{\bf{e}}_y\sin\phi+\bf{e}}_z\cos\phi)$, where $\phi$ is the wave phase and $E_{0c}=a_{c}m_e\omega_Lc/e$ is the  wave amplitude. 
The integration of exact equation of motion shows that, when this wave accelerates electrons in the ion channel to ultra-relativistic energies, they move along helical trajectories, see Fig.~\ref{fig:circular} (a), advancing forward in $x$-direction with almost speed of light and rotating along the slowly evolving circle or ellipse in $(y,z)$-plane .

Using the dimensionless variables defined by Eqs.~(\ref{eq:Eq_AA_2}), it is convenient to rewrite the equations of motion (1) and (2) in the following form:
\begin{eqnarray}
  \dot{\tilde{p}}_x =-4\mathcal{E}_c(\dot{\tilde{y}}  
\cos{\phi}+\dot{\tilde{z}} \sin{\phi}),\label{eq:curcul_EoM_x}\\
{\dot{\tilde{\bf{p}}}}_{\perp}=-
{\tilde{\bf{r}}}_{\perp}+\dot{\phi}\mathcal{E}_c
({\bf{e}}_y \cos \phi+
{\bf{e}}_z \sin \phi),\label{eq:curcul_EoM_p_perp}
\end{eqnarray}
where $\mathcal{E}_c\equiv a_{c}({\omega_p}/{\omega_L}){\tilde{v}_{ph}}{\mathcal{I}_0^{-3/2}}$. In the paraxial approximation, the time evolution of the wave phase and of the electron transverse coordinates are again determined by Eqs.~(\ref{eq:Eq_AA_3}). One can verify that the integral of motion~(\ref{eq:ENERGY_PARAX}) also holds in the case of a circularly polarized laser pulse.

We now consider an electron moving along a slowly evolving circle in the $(y,x)$-plane. In this case we can approximate transverse electron velocity as $|\dot{{\tilde{{\bf{r}}}}}_{\perp}| \approx |{{\tilde{{\bf{r}}}}}_{\perp}|\dot{\psi}$, where $\dot\psi = \omega_{\beta}/\omega_* = 1/\tilde{p}_x^{1/2}$. Making use of this relationship, we find from Eq.~(\ref{eq:ENERGY_PARAX}) that ${{\tilde{{\bf{r}}}}}_{\perp}^2 =  \dot{{\tilde{{\bf{r}}}}}_{\perp} ^2 \tilde{p}_x 
  = (1+\chi \tilde{p}_x)/2$, which implies that kinetic and potential energies associated with the transverse motion are equal to each other. Substituting the last result into the equation for the wave phase~(\ref{eq:Eq_AA_3}), we obtain that, in contrast to the case of a linearly polarized (LP) laser wave, the Doppler shifted frequency does not oscillate when ultra-relativistic electrons move 
along helical trajectories: $\dot\psi=-\omega_D/\omega_* =  -1/\tilde{p}_x-3\chi$. Thus, without applying the averaging procedure, we find that the phase shift $\theta=\psi+\phi$ satisfies the  
following equation
\begin{eqnarray}  \label{eq:theta_circular}
\dot{\theta}=\frac{\omega_{\beta}-\langle{\omega_D}\rangle}{\omega_*}=\frac{1}{\tilde{p}_x^{1/2}}-\frac{1}{\tilde{p}_x}-3\chi,
\end{eqnarray}
Substituting  $\tilde{\dot{y}}=[(1+\chi \tilde{p}_x)/2]^{1/2}  
\cos\psi\dot\psi$ and $\tilde{\dot{z}}=[(1+\chi  
\tilde{p}_x)/2\tilde{p}_x]^{1/2} \sin\psi\dot\psi$ into  
Eq.~(\ref{eq:curcul_EoM_x}), we find that
\begin{eqnarray}
  \dot{\tilde{p}}_x =-4\mathcal{E}_c\cos{\theta}[(1+\chi  
\tilde{p}_x)/2\tilde{p}_x]^{1/2},\label{eq:curcul_EoM_x}
\label{eq:p_x_circular}
\end{eqnarray}
This equation is similar to Eq.~(\ref{eq:px_SUPER}) with $\mathcal{E}_c/\sqrt{2}=\alpha \mathcal{E}$ ($\mathcal{E}_c\approx 0.5\ \mathcal{E}$).

Figure~\ref{fig:circular} (b) illustrates the similarities of electron acceleration by CP and LP laser waves with matching amplitudes. As one would expect, the phase trajectory in the case of CP laser wave looks much smoother  than in the case of a LP laser wave. 

Thus, equations~(\ref{eq:PH_TR_SUPER}) and  (\ref{eq:H_SUPER}) for the  
averaged phase trajectory and the estimate~(\ref{eq:P_MAX_SUPER}) for the maximum energy  
gain hold also for the acceleration by the CP laser wave. However, the  
initial conditions at which electrons can be actually accelerated  to  
this energy (such as pre-heating) are not the same as for the case of LP laser wave.  
Moreover  at some laser-plasma parameters, electrons can move in the  
transverse plane $y-z$ along ellipses. In this case the maximum gained  
energy can be significantly smaller than the energy obtained from  
Eq.~(22).

\subsection{Energy amplification  of pre-accelerated electrons}

Up to this point, we have assumed that  the integral of motion $\mathcal{I}_0$ is positive. For completeness, we consider the dynamics of electrons with a negative $\mathcal{I}_0$. Such a  
special case can be realized when electrons with an initially large longitudinal momentum, $p_x|_{t=0}=p_{in}\gg mc$, are accelerated by a superluminal laser wave, $v_{ph}>c$. Assuming that initially the electrons are placed on the channel axis with $p_y=p_z=0$, we obtain that 
$I_0\approx-p_{in}(v_{ph}-c)/mc^2<0$.

In contrast to the acceleration of initially cold electrons, the dynamics  of pre-accelerated electrons can be described in the paraxial approximation right from the very beginning of their motion. The process of gaining energy now has a quite regular character. Initially, the energy of transverse electron oscillations is equal exactly to zero. Then it gradually increases with time making a coupling between the wave and betatron oscillations stronger and conversion of  the wave energy to the energy of the electron longitudinal motion more  efficient. After some time, the dephasing between the wave and betatron oscillations increases and eventually  electrons start loosing their energy. A rigorous analysis of this regime is given in Appendix B. Typically, the the electron energy is amplified several times.
\section{Summary and conclusion}
In summary, we have shown that acceleration of ultrarelativistic electrons by a laser wave in an ion channel can be parametrized in the paraxial approximation using just two parameters. By averaging out betatron oscillations, we have derived reduced equations of motion that can be easily integrated to determine electron phase-space trajectories. We have found that the appearance of forbidden zones in phase-space divides electron phase-space trajectories into several unconnected segments. This results in a threshold dependence of the maximum energy gained by electrons on laser and plasma parameters. The universal scalings given by Eqs. (\ref{eq:gamma_max}) and (\ref{eq:P_MAX_SUPER}) remain the same even if electrons perform a three dimensional motion, which primarily introduces more irregularity into electron dynamics and typically delays reaching the maximum energy by the electron. The derived scalings can be used to make preliminary estimates for the contribution of DLA in wakefield accelerators. They can also be directly applied to examine DLA of electrons by a circularly polarized laser. The capability of the universal scaling theory is demonstrated by predicting electron acceleration and substantial energy gain via a previously unknown third harmonic resonance in an ion channel.


Vacuum acceleration of electrons by a laser wave is characterized by a single parameter $\gamma_*\equiv a_0^2/2\mathcal{I}_0$ ($p_{\max} = m c \gamma_*$).  Acceleration in an ion channel introduces another parameter, $\tilde{p}_* \equiv \mathcal{I}_0^2(\omega_L/2\omega_p\tilde{v}_{ph})^2$, that follows from the resonant condition that the Doppler shifted frequency is comparable to the  
frequency of betatron oscillations. Note that  $\gamma_*/\tilde{p}_*=\mathcal{E}^2$. Paradoxically, the energy gain increases in the luminal case as the the strength of the channel electric field decreases and the influence of the channel becomes weaker ($\mathcal{E}\ll 1$ and $\omega_p/\omega_L\ll 1$). The downside of working in this limit is that electrons must be initially significantly preheated in order to attain the predicted maximum energy. In the super-luminal case, one more parameter, characterizing the effective wave dispersion, comes into play: $\chi\propto  
(v_{ph}/c-1)/(\omega_p^2/\omega_L^2)$. It increases the Doppler shifted  
frequency and introduces an additional longitudinal momentum scale, $p_+ \propto  
p_*/\chi^2$, which allows electrons to remain in resonance at significantly higher values of electron momentum.

Our model relies on a number of key simplifications whose applicability must be examined in the context of a specific problem of interest. In the developed model, we assume a fully evacuated ion channel and a plane wave, neglecting longitudinal electric fields of the laser wave. The transverse variation of the laser field might however be important in the case of a tightly focused laser pulse in an ion channel. The channel also might not be fully evacuated and thus it might contain residual electrons that are heated by the laser wave. It is also important to point out that the phase velocity is not an independent parameter and it is strongly influenced by the channel radius and the plasma density distribution. The impact of these aspects on electron acceleration must be examined via self-consistent first-principle PIC simulations. Nevertheless, such a simplified description  of  acceleration of electrons by the laser wave has a powerful predictive capability demonstrated in Refs.~\cite{zhang_prl,zhang_ppcf}.
 
This work was motivated by laser-plasma interactions at sub-critical plasma densities, but  the developed approach is much more general and it is not limited just to the regime where $\omega_p \ll \omega$. The theory is well suited to make meaningful predictions regarding electron acceleration in near-critical and over-critical plasmas, provided that such plasmas are relativistically transparent to the incoming high-intensity laser pulse.  Wave propagation in a relativistically transparent plasma is a critical aspect of this regime that should be addressed self-consistently.

This work was supported by DOE grants DE-SC0007889 and DE-SC0010622, and by an AFOSR grant FA9550-14-1-0045.
AVA was supported by the U.S. Department of Energy [National Nuclear Security Administration] under Award Number DE-NA0002723, by AFOSR Contract No. FA9550-14-1-0045, U.S. Department of Energy - National Nuclear Security Administration Cooperative Agreement No. DE-NA0002008, and U.S. Department of Energy Contract No. DE- FG02- 04ER54742. 
 \appendix
\section{Averaging procedure for Lorentz force.}
Let us consider a luminal case with $\chi=0$ and assume that the longitudinal electron momentum changes slowly with time  and calculate. Near the resonance, $\langle{\omega}_D\rangle=l\omega_{\beta}$, the particle oscillations' phase and the wave phase satisfy equations:
$\dot{\psi} =\omega_{\beta}$ and $\dot{\phi} =-2l\omega_{\beta}{\sin^2\psi}$, so that 
\begin{eqnarray}
\phi= 
\langle\theta\rangle - l\psi+\frac{l}{2}\sin {2 \psi},\label{eq:SM1}
\end{eqnarray} 
where $\theta=l\psi+\phi$, and averaging is performed over betatron phase $\psi$. Using this relationship, we can average the oscillating factor in the Lorentz force 
\begin{eqnarray}
\begin{aligned}
 \langle \cos\phi\cos\psi\rangle=\frac{1}{{2 \pi}}\int_{0}^{2 \pi}\sin\psi\cos\varphi d\psi=
\alpha_l\cos\langle\theta\rangle,\label{eq:SM2}
\end{aligned}
\end{eqnarray} 
where
\begin{eqnarray}
\begin{aligned}
\alpha_l= \frac{1}{{2 \pi}}\int_{0}^{2 \pi}\sin\psi\sin\Big(l\psi -\frac{l}{2}\sin {2 \psi}\Big)d\psi=\\
\frac{1}{2}[1-(-1)^l](-1)^{\frac{l+3}{2}}\Big[J_{l_1}\Big(\frac{l}{2}\Big)-J_{l_2}\Big(\frac{l}{2}\Big)\Big].\label{eq:SM3}
\end{aligned}
\end{eqnarray} 
where $l_{1,2}=(l \mp 1)/2$. Thus, $\alpha_1=0.348101$, $\alpha_2=0$, $\alpha_3=-0.162924$,  $\alpha_4=0$, and $\alpha_5=0.114729$. In the superluminal case, these coefficients depend on the parameters $\chi$. However,  this dependence can be neglected at small $\chi$.  

\section{Acceleration  initially pre-accelerated electrons}

We consider an  initially pre-accelerated electron with $p_x|_{t=0}=p_{in}>>mc$ and  $p_y|_{t=0}=0$ placed on the channel axis so that
 \begin{eqnarray}
 \mathcal{I}_0\approx-(v_{ph}/c-1)p_{in}/mc.  \label{eq:Integr}
\end{eqnarray} 
Transverse electron energy  grows with  the longitudinal momentum as  
$\varepsilon_{\perp}= -|I_0| m_e c^2+p_x(v_{ph}-c)$. 
\begin{figure}[t]
 \centering
		\quad \includegraphics[height=0.110\textheight,width=0.85\columnwidth]{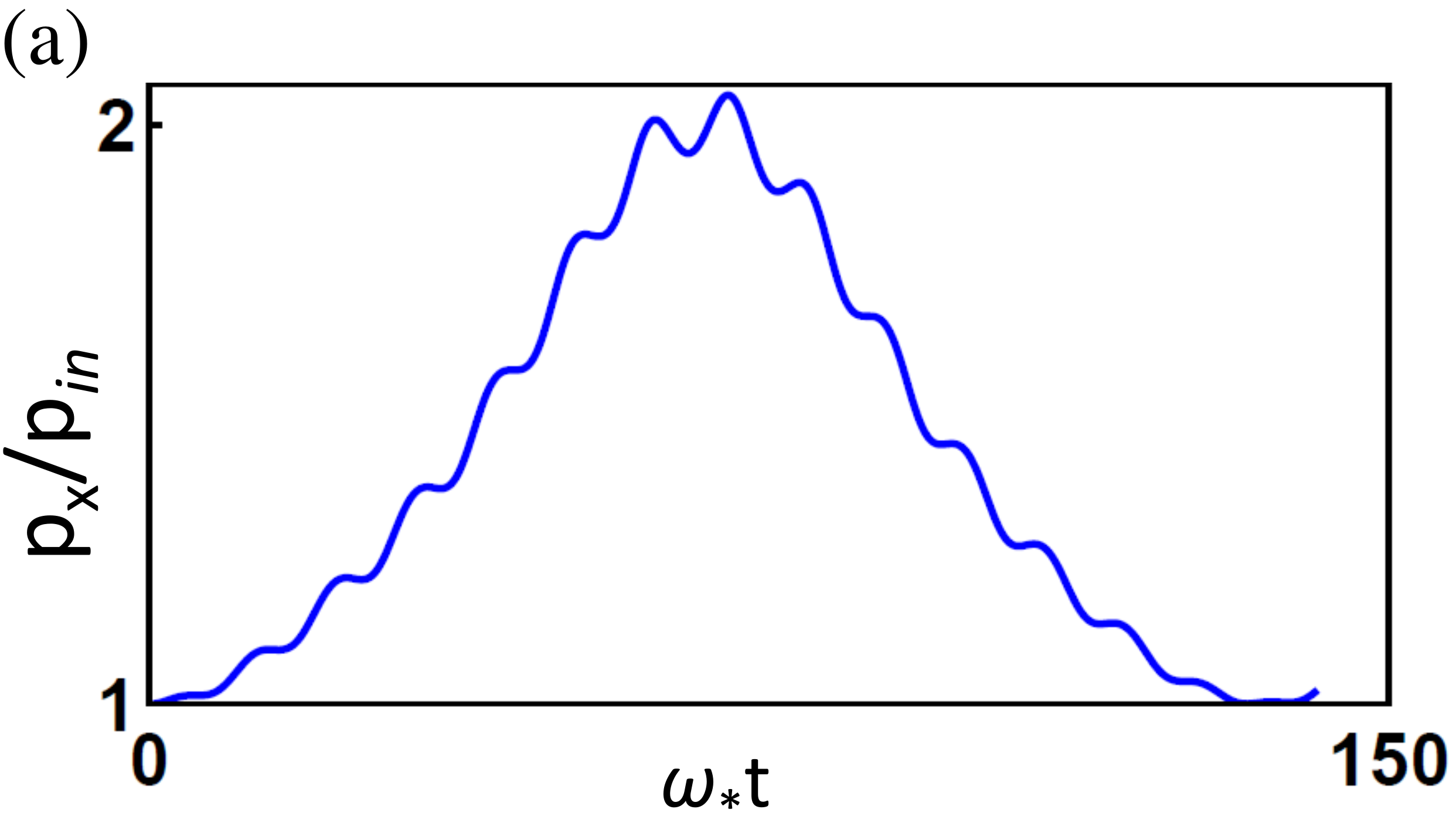}\\
		\includegraphics[height=0.13\textheight,width=0.85\columnwidth]{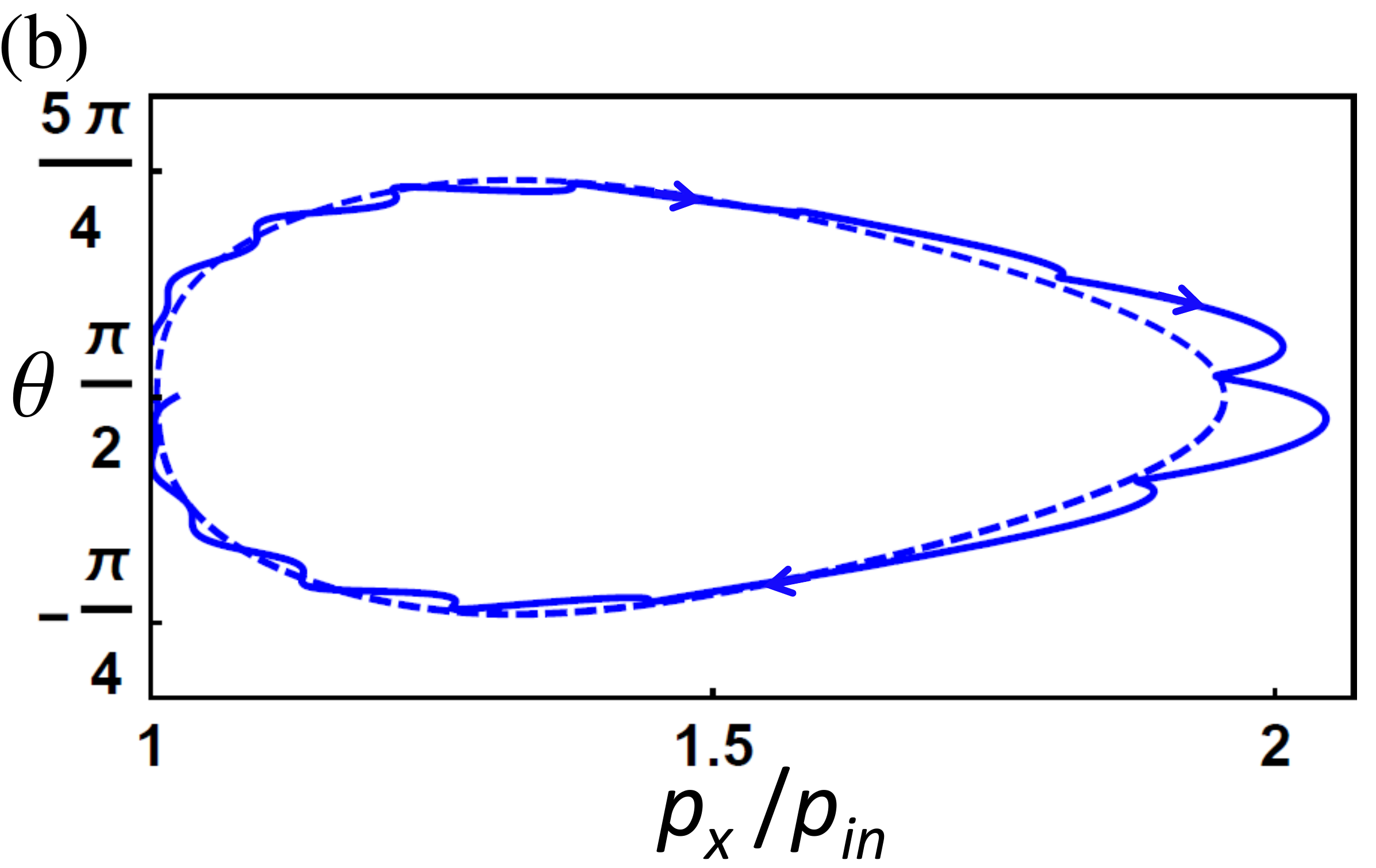}
			\caption{(a)  Dependence $p_x$ on $t$  at $|\chi|=0.147$ and $|\mathcal{E}|=0.368$ ($a_0=6$, $\omega_p/\omega_L =\sqrt{0.1}$, $n_e/n_{cr}=0.1$,  $p_{in}=300m_ec$, $\mathcal{I}_0=-3$ and $v_{ph}/c=1.01c$). (b)~Trajectory  in the phase space (${\theta},p_x$) from exact (solid line) and  averaged  (dashed line) equations of motion.  } \label{fig:phase_space_negative_I0}
\end{figure}
In the paraxial approximation, the equations for the longitudinal momentum and the wave phase take the following form
\begin{eqnarray}
 \dot {\tilde{p}}_x =-4|\mathcal{E}|\dot{\tilde{y}} \cos{\phi},\label{eq:B_EqOfM1}\\
\dot{\phi} =-2(\dot{\tilde{y}}^2+|\chi|),\label{eq:B_EqOfM2}, 
\end{eqnarray} 
and the integral of motion is given by 
\begin{eqnarray}
 \tilde{p}_x\dot{\tilde{y}}^2+\tilde{y}^2=-1+|\chi|\tilde{p}_x,  \label{eq:EqOfM}
\end{eqnarray} 
where
\begin{eqnarray}
|\mathcal{E}|= a_0(\omega_p/\omega_L)\tilde{v}_{ph}	/|\mathcal{I}_0|^{3/2}, \\
|\mathcal{\chi}|=(|\mathcal{I}_0|/\tilde{v}_{ph}^2)(v_{ph}/c-1)/(2\omega_p^2/\omega_L^2), \label{eq:NORMALIZ1}\\
 \tilde{p}_x=p_x/p_*=(p_x/m_ec)/(I_0^2/\tilde{v}_{ph}^2)(\omega_L^2/2\omega_p^2),\label{eq:NORMALIZ2}\\ 
\tilde{y}=k_py/2\sqrt{|\mathcal{I}_0|}, \quad\tilde{t}=\omega_*t=t\tilde{v}_{ph}\omega_p^2/(\omega_L|\mathcal{I}_0|).
  \end{eqnarray} 
and $\mathcal{I}_0$ is determined by Eq.~(\ref{eq:Integr}). It follows from Eqs.~(\ref{eq:NORMALIZ1}) and (\ref{eq:NORMALIZ2}) that the initial momentum in dimensionless variables is $\tilde{p}_{in}=1/|\chi|$
and that the initial  energy of the transverse oscillations is equal to zero. The laser wave slowly pumps energy into the transverse oscillations which is converted to the increasing momentum/energy of the longitudinal motion, see  Fig.~\ref{fig:phase_space_negative_I0} (a) according to the relationship: $\epsilon_{\perp}/m_ec^2 =|\mathcal{I}_0|(p_x/p_{in}-1)$.

Assuming that the longitudinal momentum changes slowly, we obtain $\tilde{y}={(-1+ |\chi|\tilde{p}_x)^{1/2}}\sin\psi$, $\dot{\psi}\approx 1/{\tilde{p}_x^{1/2}}$, $\dot{\tilde{y}}\approx {[(-1+ |\chi|\tilde{p}_x)/\tilde{p}_x]^{1/2}}\cos\psi$ and $\langle\dot{\phi}\rangle=1/\tilde{p}_x -3|\chi|$. The phase shift between the laser wave and betatron oscillations satisfies the following equation:
\begin{eqnarray}
\langle\dot{\theta}\rangle=1/{\tilde{p}_x^{1/2}}+1/\tilde{p}_x -3|\chi|.
 \label{eq:APP_phase_shift}
\end{eqnarray}
Near the resonance $\omega_{\beta}\approx \langle\omega_D\rangle$  the
equation for the longitudinal momentum takes the following form
\begin{eqnarray}
 \dot{\tilde{p}}_x=-4\alpha \mathcal{E}\cos{ \langle{\theta}\rangle}[(-1+|\chi| \tilde{p}_x)/\tilde{p}_x]^{1/2}.
 \label{eq:APP_px_SUPER}
\end{eqnarray}
 After eliminating the explicit dependence on time by dividing Eq.~(\ref{eq:APP_phase_shift}) by Eq.~(\ref{eq:APP_px_SUPER}) we find that
\begin{eqnarray}
\frac{d }{d \tilde{p}_x}\sin \langle\theta\rangle= -\frac{1+1/\tilde{p}_x^{1/2} -{3}|\chi| \tilde{p}_x^{1/2}}{4\alpha |\mathcal{E}| \sqrt{(-1+ |\chi|\tilde{p}_x)}}
 \label{eq:AP_EQ_PH_TR_SUPER}
\end{eqnarray}
The resulting phase trajectory is given by  
\begin{eqnarray}
\sin \theta=-\frac{1}{4\alpha |\mathcal{E}|}
H({|\chi|,|\chi|\tilde{p}_x})+C,\label{eq:AP_TRAJECTORY1}\\
H(z,\chi)\equiv \frac{3\sqrt{z}\mu(z)+\ln[\sqrt{z}+\mu(z)]}{\sqrt{|\chi|}}-\frac{2\mu(z)}{|\chi|}.\label{eq:AP_TRAJECTORY2}
\end{eqnarray}
where $\mu(z)\equiv \sqrt{z-1}$.

\begin{figure}[t]
 \centering
 	  \includegraphics[height=0.150\textheight,width=0.95\columnwidth]{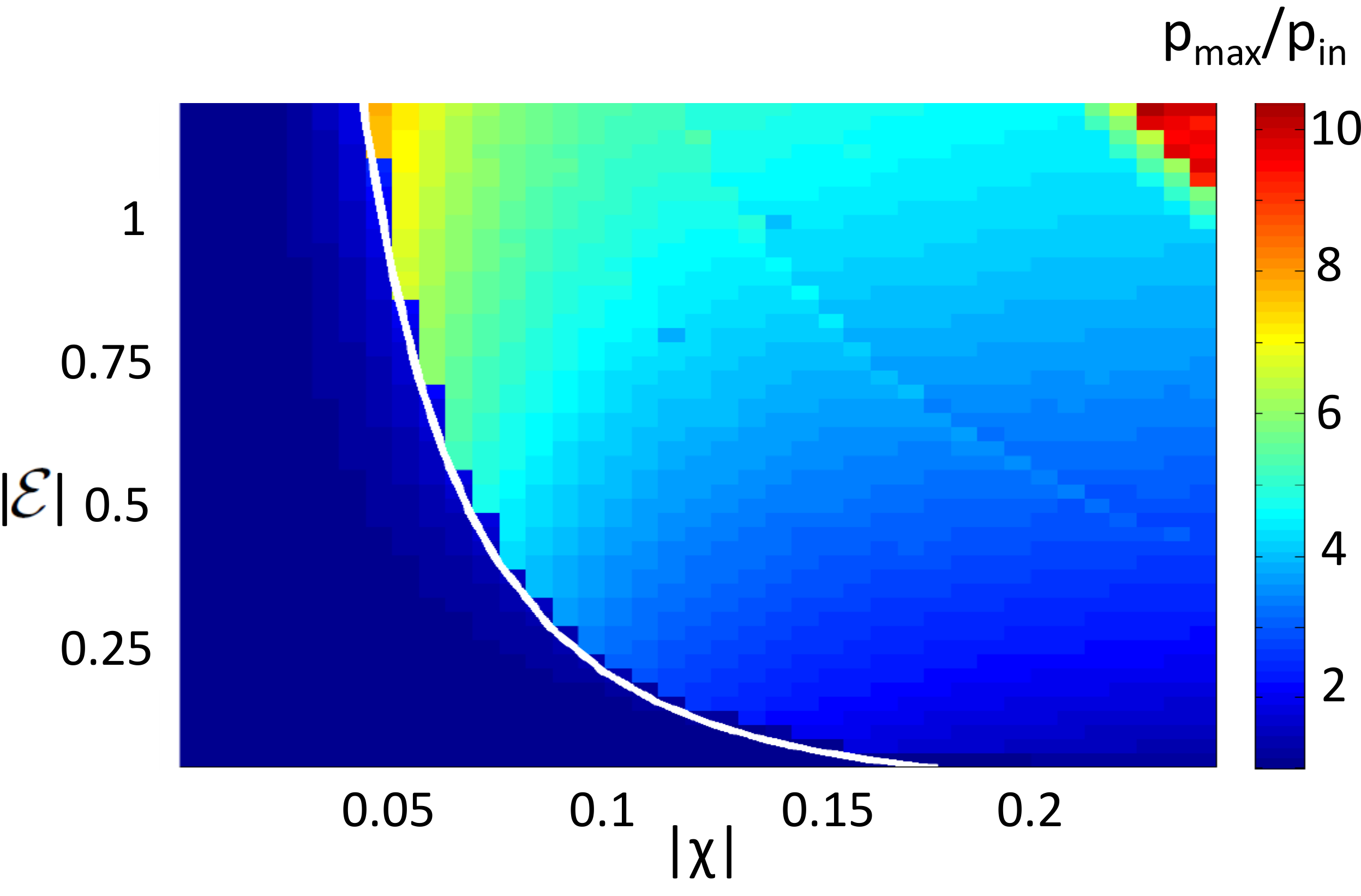}
			\caption{  Amplification factor $p_{\max}/p_{in}=A_F(|\chi|,|\mathcal{E}|)$ for initially pre-accelerated electrons.} \label{fig:phase_space_negative_I1}
\end{figure}

The exact resonant condition $\langle{\omega}_D\rangle=\omega_{\beta}$ is satisfied at only one point: $\tilde{p}_{+}=(1/6\chi)^2(1+\sqrt{1+12|\chi|})^2$. Since Eq.~(\ref{eq:APP_px_SUPER}) is valid only near the resonance, this point should be passed by the electron during its acceleration: $\tilde{p}_{+}>\tilde{p}_{in}$ and hence the parameter $|\chi|$ must be less than $1/4$.   
As one can see from Eqs.~(\ref{eq:AP_TRAJECTORY1}) and (\ref{eq:AP_TRAJECTORY2}), the energy amplification depends only on parameters $|\mathcal{E}|$ and $|\chi|$:
\begin{eqnarray}
p_{\max}/p_{in}=|\chi|p_{\max}=A_F(|\chi|,|\mathcal{E}|)
\label{eq:AP_max}
\end{eqnarray}
where function $A_F$ is depicted in Fig.~\ref{fig:phase_space_negative_I1}. There is  a threshold (white line) in the parameter space $(|\chi|,|\mathcal{E}|)$  associated with the minimum of the function $H(|\chi|,|\chi|\tilde{p}_x)$ which is reached at the resonant point $\tilde{p}_x=\tilde{p}_+$. When parameters  $|\chi|$ and $|\mathcal{E}|$ belong to  the area at the left of the white line, electrons move along small segment of the averaged phase trajectory and their longitudinal momentum $\tilde{p}_x$ cannot exceed $\tilde{p}_+$. On contrary, when parameters  $|\chi|$ and $|\mathcal{E}|$ belong to  the area at the right of the white line, electrons move along the entire phase trajectory   reaching  larger values of the momentum. 
The threshold can be determined from the following condition 
\begin{eqnarray}
 \frac{1}{4|\mathcal{E}|}|H(|\chi|,{|\chi|\tilde{p}_+})-H(|\chi|,{|\chi|\tilde{p}_{in}})|\approx C_0.\label{eq:AP_MAX}
\end{eqnarray}
 The numerical analysis shows that the constant $C_0$ in the right-hand side is equal to 3 rather than the theoretical value 2.

It is interesting to note that there is a red area in the right upper corner in Fig.~\ref{fig:phase_space_negative_I0} (a) with large amplification factors. The parameters in this area correspond to third harmonic (or some mixture of harmonics) resonances. 


  \nocite{*}

 

 \end{document}